\documentclass[12pt]{article}
\usepackage{authblk}
\usepackage{amsmath,amssymb}
\usepackage{setspace,subfig}
\usepackage{natbib}
\usepackage{longtable}
\usepackage{relsize}
\usepackage[hidelinks]{hyperref}
\renewcommand{\harvardurl}[1]{\textbf{URL:} \url{#1}}
\usepackage{booktabs}
\usepackage{pgf}
\usepackage{multirow}
\usepackage{setspace}
\usepackage{times}
\usepackage{mathtools}
\usepackage{tikz}
\usepackage{times}
\usetikzlibrary{arrows,shapes.arrows,shapes.geometric,patterns,
	shapes.multipart,backgrounds,decorations.pathmorphing,positioning,fit,automata}
\tikzset{
	>=stealth',
	true/.style={
		rectangle,
		draw=black, very thick,
		text width=6.5em,
		minimum height=2em,
		text centered,
		fill=gray, opacity = 0.5},
	punkt/.style={
		rectangle,
		rounded corners,
		draw=black, very thick,
		text width=6.5em,
		minimum height=2em,
		text centered},
	shade/.style={
		circle,
		draw=black, very thick, fill=gray!50,
		text centered}
}
\usepackage{subfig}
\usepackage{enumerate}

\newtheorem{prop}{Proposition}
\usepackage{booktabs}
\usepackage{amsfonts}
\usepackage[latin1]{inputenc}
\usepackage{soul,color}
\usepackage{chngcntr}
\usepackage{verbatim}
\usepackage{changepage}
\usepackage{amssymb}
\usepackage{pgf}
\newcommand{\bm}[1]{\mbox{\boldmath{$#1$}}}

\usepackage{caption} 
\usepackage[left=1in,right=1in,top=1in,bottom=1in]{geometry}

\usepackage{xfrac}
\usepackage{graphicx}
\newcommand{\ind}{\rotatebox[origin=c]{90}{$\models$}}

\newtheorem{theorem}{Theorem}

\usepackage[T1]{fontenc}
\newtheorem{assumption}{Assumption}

\newtheorem{proof}{Proof}
\allowdisplaybreaks
\newcommand\numberthis{\addtocounter{equation}{1}\tag{\theequation}}

\usepackage{algorithm}
\date{}
\usepackage{tikz}

\usepackage{soulpos}
\soulregister\cite7
\soulregister\citep7
\soulregister\ref7
\soulregister\emph7
\soulregister\eqref7
\soulregister\pageref7
\definecolor{mygreen}{RGB}{144,241,47}

\newcommand{\logit}{\text{\it logit}}

\def\bSig\mathbf{\Sigma}

\usepackage[figuresright]{rotating}

	{\raise0.5ex\hbox{$#1$}\! \left/ \! \lower0.5ex\hbox{$#2$}\right.}
	
\begin{document}

		\markboth{L. Wang, X. Meng, T.S. Richardson and J.M. Robins}{Coherent modeling of longitudinal causal effects}
	

 \centerline{\large\bf Coherent  modeling of longitudinal causal effects on binary outcomes}
\vspace{.25cm}
\vspace{.4cm} \centerline{$^{1}$Linbo Wang, $^{2}$Xiang Meng, $^{3}$Thomas S. Richardson, $^{4}$James M. Robins} \vspace{.4cm} \centerline{\it $^{1}$University of Toronto,
$^{2}$Harvard University,
$^{3}$University of Washington, $^{4}$Harvard T.H. Chan School of Public Health
} \vspace{.55cm} 
\par

\begin{abstract}
	Analyses of biomedical studies often necessitate modeling longitudinal causal effects. The current focus on personalized medicine and effect heterogeneity makes this task even more challenging. Towards this end, structural nested mean models (SNMMs) are fundamental tools for studying heterogeneous treatment effects in longitudinal studies.   However, when outcomes are binary, current methods for estimating multiplicative and additive SNMM parameters suffer from variation dependence between the causal parameters and the non-causal nuisance parameters. This leads to a series of difficulties in interpretation, estimation and computation. These difficulties have hindered the uptake of SNMMs in biomedical practice, where binary outcomes are very common. We solve the variation dependence problem for the binary multiplicative SNMM via a reparametrization of the non-causal nuisance parameters.  Our novel nuisance parameters are variation independent of the causal parameters, and hence allow for coherent modeling of heterogeneous effects from longitudinal studies with binary outcomes. Our parametrization also provides a key building block for flexible doubly robust estimation of the causal parameters. Along the way, we prove that an additive SNMM with binary outcomes does not admit a variation independent parametrization, thereby justifying the restriction to multiplicative SNMMs.  
\end{abstract}



\section{Introduction}
\label{sec:intro}

\begin{figure}[!h]
	\centering
	\begin{subfloat} 
		\centering
	\scalebox{0.8}{
		\begin{tikzpicture}[->,>=stealth',shorten >=1pt,auto,node distance=2.5cm, shape=ellipse,
		semithick, scale=0.3]
		pre/.style={->,>=stealth,semithick,blue,line width = 1pt}]
		\tikzstyle{every state}=[fill=none,draw=black,text=black, shape=ellipse]
		\node[state] (A0)             {$A_0$};
		\node[state] (L) [right of=A0] {$L_1$};
		\node[state]         (A1) [right of=L] {$A_1$};
		\node[state](Y) [right of = A1]{$Y$};
		\node[shade](U) [below = 1cm of A1]{$U$};
		\node[]() [below  = 2cm of L] {(a): Full DAG};
		\path
		(A0) edge node {} (L)
		(L) edge node {} (A1)
		(A1) edge node {} (Y)
		(U) edge node {} (L)
		(U) edge node {} (Y);
		\draw[->, bend left] (A0) edge (A1);
		\draw[->, bend left] (L) edge (Y);
		\draw[->, bend left] (A0) edge (Y);
		\end{tikzpicture}
		}
	\end{subfloat}
	\kern20pt
\begin{subfloat} 
\centering
	\scalebox{0.8}{
		\begin{tikzpicture}[->,>=stealth',shorten >=1pt,auto,node distance=2.5cm, shape=ellipse,
		semithick, scale=0.3]
		pre/.style={->,>=stealth,semithick,blue,line width = 1pt}]
		\tikzstyle{every state}=[fill=none,draw=black,text=black, shape=ellipse]
		\node[state] (A0)             {$A_0$};
		\node[state] (L) [right of=A0] {$L_1$};
		\node[state]         (A1) [right of=L] {$A_1$};
		\node[state](Y) [right of = A1]{$Y$};
		\node[shade](U) [below = 1cm of A1]{$U$};
			\node[]() [below  = 2cm of A1] {(b):  DAG under the g-null};
		\path
		(A0) edge node {} (L)
		(L) edge node {} (A1)
		(U) edge node {} (L)
		(U) edge node {} (Y);
		\draw[->, bend left] (A0) edge (A1);
		\end{tikzpicture}
		}
	\end{subfloat}
	\caption{DAGs illustrating time-varying treatments and confounders. The baseline covariates $L_0$ are omitted for brevity. Variables $A_0,L_1,A_1,Y$ are observed; $U$ is unobserved.}
	\label{fig:basic}
\end{figure}
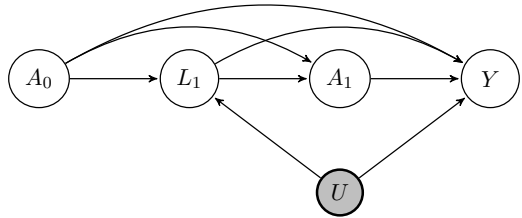

In biomedical studies researchers are often interested in inferring causal effects from longitudinal studies with time-dependent exposures. For example, suppose one is interested in estimating the (joint) effect of maternal stress on childhood illness from longitudinal observational data. 
The relationships among observed variables may be represented by the causal directed acyclic graph \citep[DAG,][]{pearl2009causality} in Figure \ref{fig:basic}(a), in which  $A_0$ and $A_1$ denote maternal stress levels at baseline  and the first follow-up, respectively,  $L_1$ is  the intermediate covariate encoding whether or not the child is ill at the first follow-up, and $Y$ is the outcome of interest encoding whether or not the child is ill at the second follow-up. The node $U$ denotes unmeasured variables such as the child's underlying immune status.  There may also be covariates  $L_0$ measured at baseline, in which case one can add $L_0$ and a directed edge from $L_0$ to every other node in Figure \ref{fig:basic}(a).  

As pointed out by \cite{robins1986new},
conventional regression adjustment methods cannot be used to estimate the joint effect of $A_0$ and $A_1$,  regardless of whether or not one adjusts for the time-dependent confounder $L_1$ in the regression. Instead, \cite{robins1986new} proposed the so-called g-formula  
\begin{equation}
\label{eqn:g-formula2}
    E[Y(a_0,a_1)] = \sum\limits_{l_0, l_1} E[Y\mid l_0,a_0, l_1, a_1] f(l_1\mid l_0, a_0) f(l_0),
\end{equation}
where $Y(a_0,a_1)$ denotes the potential outcome, $E[Y\mid l_0,a_0, l_1, a_1] = E[Y\mid L_0 = l_0, A_0=a_0, L_1=l_1, A_1=a_1],$ $f(l_1\mid l_0, a_0)$  and $f(l_0)$ represent the (conditional) density of $L_1$ and $L_0$, respectively. Application of the g-formula in practice, however, is subject to the g-null paradox {\citep{robins1986new,robins1997estimation}}. Specifically, under the causal null hypothesis represented by Figure \ref{fig:basic}(b), in general, both $E[Y\mid l_0,a_0, l_1, a_1]$ and $f(l_1\mid l_0, a_0)$ depend on $a_0$.  {In this case, it is very challenging to specify models for 
$E[Y\mid l_0,a_0, l_1, a_1]$ and $f(l_1\mid l_0, a_0)$ that allow for each term to depend on $a_0$ while at the same time being compatible with the causal null hypothesis that \eqref{eqn:g-formula2} does not depend on $a_0$.} 


In response to the g-null paradox, \cite{robins1994correcting} and \cite{robins2000marginal} introduced  structural nested mean models (SNMMs) and  marginal structural models (MSMs), respectively. {Compared to the g-formula, both of them may be used to impose parsimonious models for heterogeneous treatment effects.}
However,  while SNMMs
model how the effect of treatment at each time point is modified by the entire past covariate and treatment history, MSMs are only able to model effect modification by covariates measured at start of follow up. Furthermore, a particular implementation of SNMMs estimates the optimal treatment strategy among all possible strategies (that depend on the observed data) through a semiparametric version of  dynamic programming, often referred to as A-learning in the dynamic treatment regime literature. This is in contrast to dynamic MSMs that  can only be used to estimate the optimal strategy among a (smaller) prespecified class of regimes \citep{murphy2003optimal,robins2004optimal,henderson2010regret,schulte2014q,shi2018high,qian2021estimating}. Consequently, SNMMs
have gained popularity among practitioners in recent years because of the growing interest in discovering effect heterogeneity.



In the simplest  case where there is only one follow-up,
the SNMM  is known as the structural mean model (SMM) and takes the following form:
\begin{equation}
\label{eqn:smm}
g(E[Y(1)\mid A_0=1, L_0=l_0]) - g(E[Y(0)\mid A_0=1, L_0=l_0]) =  B(l_0; \alpha),
\end{equation}
where $g$ is the link function, $Y(a_0), a_0=0,1$ denotes the potential outcome  and $B(l_0;\alpha)$ is a function known up to a finite-dimensional parameter $\alpha$ such that $B(l_0;0) = 0$. A leading special case is the linear specification $B(l_0;\alpha) = \alpha^T l_0.$
Under the sequential ignorability assumption \citep{robins1986new}, the structural model  \eqref{eqn:smm} and the linear specification for $B(l_0;\alpha)$ imply the following observed data model:  
\begin{equation}
\label{eqn:smm_obs}
g(E[Y\mid A_0 = 1, L_0=l_0]) - g(E[Y\mid A_0=0, L_0=l_0]) = \alpha^T l_0.
\end{equation} 
Model \eqref{eqn:smm_obs} is semiparametric as it does not specify  the full regression function $g(E[Y \mid  A_0=a_0, L_0=l_0])$. To enable maximum likelihood estimation, one may assume an additional baseline mean model, $E[Y\mid A_0=0, L_0=l_0; \zeta]$, resulting in a generalized linear model (GLM). Alternatively, 
with the log or identity link, estimation of $\alpha$  is often  based on doubly robust g-estimation methods.

	
	An outstanding problem in the application of SNMMs to realistic setting is that when the outcome $Y$ is binary and the link $g$ is the log or the identity function, even for the simple point exposure case, a baseline mean model $P(Y=1\mid A_0=0, L_0 = l_0; \zeta)$ {is variation dependent on, and hence can be incompatible with, the SMM}.
	In this case, maximum likelihood estimation requires constrained optimization in a restricted parameter space, and  with a new covariate value for $L_0$, the maximum likelihood estimator (MLE) $(\widehat{\alpha}_{\text{mle}}, \widehat{\zeta}_\text{mle})$ may still imply a fitted risk  $\widehat{P}(Y=1\mid A_0=1, L_0=l_0)$ to be greater than one. Additionally, the g-estimators fail to be ``truly doubly robust'' \citep{wang2021estimation} because of the incompatible baseline model.
	On the other hand, when the link $g$ is the logistic function, it is not possible to use g-estimation methods for estimating parameters in SNMMs \citep{robins2000marginal2}. In particular, unlike the case with the additive or multiplicative SNMM, it is not possible to guarantee consistent estimation of logistic SNMM parameters even in randomized trials \citep{robins2004estimation}.  Inference for logistic SNMMs is also considerably more complicated than that for the multiplicative or additive SNMMs; see, for example, \cite{vansteelandt2003causal,robins2004estimation} and \cite{matsouaka2014likelihood}. Furthermore, the logistic SNMMs estimate odds ratios which are not collapsible \citep{rothman2008modern}. The non-collapsibility  of  odds ratios  limits the interpretability and generalizability of estimates from logistic SNMMs. 
	
	For the reasons mentioned above, over the past two decades SNMMs were regarded as inappropriate for inferring causal effects with binary outcomes \citep[e.g.][]{robins2000marginal2,daniel2013methods,vansteelandt2010estimation,vansteelandt2014structural}. 
	\cite{richardson2017modeling} offered a novel approach to overcoming these problems in the point exposure case, with a binary treatment and binary outcome. The key observation is that the baseline risk model included in a GLM is often not of primary interest; instead, it is a \emph{nuisance} model to aid estimation of the SMM parameter. To resolve the variation dependence between the \emph{conventional} nuisance model and the SMM, they introduce a novel nuisance model that is variation independent of the SMM with the log or identity link. In conjunction with the SMM, their nuisance model gives rise to a likelihood for $P(Y=1\mid A_0=a_0, L_0=l_0)$ so that one can use unconstrained maximum likelihood for estimation. Furthermore, it permits true doubly robust g-estimation as
	the nuisance model is compatible with the SMM. In a recent paper, \cite{yin2022multiplicative}   extended their method to accommodate a categorical exposure or, under an additional monotonicity assumption, a continuous exposure.
	
	In this paper, we study the more challenging case of time-varying treatments. To focus on the main ideas, we primarily discuss the special case in which the time-varying treatments, time-varying confounders and outcome are all binary; note that the baseline confounders are still allowed to be continuous or discrete. {This special case is prevalent in modern applications such as micro-randomized trials considered by \cite{qian2021estimating}, where the time-varying confounder denotes the availability of a participant for the treatment. We also discuss extensions to accommodate categorical or (under additional assumptions) continuous time-varying treatments and confounders in Section \ref{sec:extension-general-treatment} of the Supporting Information.}

	Specifically,  we show that  unlike in the point exposure case,  the causal  parameters of binary additive SNMMs are generally variation \emph{dependent} on each other. In comparison, the causal  parameters of binary multiplicative SNMMs  are  variation \emph{independent} of each other. For the latter, in parallel to the work of \cite{richardson2017modeling}, we develop novel nuisance models that are variation independent of the multiplicative SNMMs. 	
	{This is more challenging than in the point exposure problem as we need to ensure compatibility among a much larger set of models: the number of these models grows exponentially with the number of time points considered.}
	
	The rest of this article is organized as follows. In Section \ref{sec:framework} we review the work of \cite{richardson2017modeling} on the point exposure case, as well as that of \cite{robins1994correcting} on SNMMs. In Section \ref{sec:parameterization} we present our main results on parameterizations for the binary multiplicative and additive SNMMs. 
	We summarize our estimation approaches in Section \ref{sec:estimation}, and
 then illustrate our approach via simulations and an application to the Mothers' Stress--Children's Morbidity Study in Sections \ref{sec:data} and \ref{sec:real_data}, respectively. We end with a discussion in Section \ref{sec:discussion}.

	\section{Framework and problem description}
	\label{sec:framework}
	
	\subsection{Review of coherent models for the relative risk and risk difference}
	\label{sec:smm}
	
	Consider a biomedical study with binary treatment $A_0$,  outcome $Y$ and general baseline covariates $L_0$.
	Under the conditional ignorability condition that $A_0\; \ind\; Y(a_0)\mid L_0,$ the SMM \eqref{eqn:smm} is equivalent to the observed data model \eqref{eqn:smm_obs}. With a continuous $Y$, a common approach to estimate the SMM parameter $\alpha$ is to assume a GLM, 
	\begin{equation}
	    \label{eqn:glm}
	    g(E[Y\mid A_0, L_0]) = \zeta^T L_0 + (\alpha^T L_0)A_0
	\end{equation} which is equivalent to assuming model \eqref{eqn:smm_obs} and an additional nuisance model,
	\begin{equation}
	\label{eqn:baseline}
	    g(E[Y\mid A_0 = 0, L_0]) = \zeta^T L_0.
	\end{equation}
The GLM \eqref{eqn:glm} with $g(x)\!=\!x$ and $g(x)\!=\!\log(x)$,
specify, respectively, linear and log-linear models for the mean. 
	Inference for $\alpha$ (and $\zeta$) may then be based on the likelihood function $L(Y\mid A_0, L_0; \alpha, \zeta).$

	Now consider the case where $Y$ is binary; the left hand side of
	\eqref{eqn:smm_obs} is known as the conditional risk difference $RD(l_0)$ when $g(x) = x$, and the conditional log relative risk $\log(RR(l_0))$ when $g(x) = \log(x)$.  Linear or log-linear regression is often regarded inappropriate for binary $Y$ because the nuisance model \eqref{eqn:baseline} is variation dependent on the model of interest \eqref{eqn:smm_obs}, in the sense that the range of $\alpha$ depends on the specific value of $\zeta$. For example, suppose that the baseline risk $E[Y\mid A_0=0, L_0=l_0] = 0.5,$ then $RD(l_0) \in [-0.5,0.5]$ and $RR(l_0) \leq 2.$ The variation dependence has led to a series of problems in interpretation, estimation and computation. As a result, the multiplicative and additive SMMs have been considered inappropriate for dealing with a binary outcome \citep[e.g.][]{daniel2013methods}.

		To solve this problem, 	
	\cite{richardson2017modeling} introduce a novel nuisance function, the $l_0$-specific odds product: $
	OP(l_0) =  {p_0(l_0) p_1(l_0)}/\{(1-p_0(l_0))(1-p_1(l_0))\}
	$
	and show that it is variation independent of both the $RD(l_0)$ and $\log(RR(l_0))$; here $p_{a_0}(l_0) = P(Y=1\mid A_0=a_0, L_0=l_0).$ Furthermore, given $l_0$,  the mappings 	
	\begin{equation}
	    \label{eqn:map}
	    	(RD(l_0),  OP(l_0))  \rightarrow (p_1(l_0), p_0(l_0)) \text{\quad and\quad }
	(\log RR(l_0),  OP(l_0)) \rightarrow (p_1(l_0), p_0(l_0))
	\end{equation}
	are both smooth bijections from $\mathbb{D} \times \mathbb{R}^{+}$ to $(0,1)^2$, where $\mathbb{R}^+ = (0,\infty)$, $\mathbb{D} = (-1,1)$ for $RD(l_0)$ and $\mathbb{D} = \mathbb{R}$ for $\log(RR(l_0))$.
	Figure \ref{fig:one} in the Supporting Information gives an illustration. One can see that  each contour line of relative risk or risk difference intersects with a given contour line of the odds product at one and only one point, so that these parameters are variation independent \emph{and} the maps in \eqref{eqn:map} are bijections. The latter feature is important as it allows for likelihood inference and risk predictions with these models.

	\subsection{Structural nested mean models: Introduction}
	\label{sec:snmm}
	
	In this paper we consider a more general biomedical study with longitudinal measurements at multiple time points $0, \ldots, K$. Let $L_k; k=1,\ldots, K$ and $A_k; k=0, \ldots, K$ denote the binary covariate measurement and binary treatment indicator at time $k$, respectively, and let $Y$ be the outcome measured at time $K+1$; recall that $L_0$ denotes general baseline covariates that can be possibly high-dimensional and are not necessarily binary. The assumption on binary $A_k$ and $L_k$ may be relaxed; see Section \ref{sec:extension-general-treatment} in the Supporting Information. We also let $\overline{A}_k$ and $\overline{L}_k$ denote the treatment and covariate history up to time $k$, that is $\overline{A}_k = (A_0, \ldots, A_k)$ and $\overline{L}_k = (L_0,\ldots, L_k)$. We presume that the covariate measurement  precedes treatment at the same time point. We use $Y(\overline{a}_k,\bm 0)$ to denote the outcome that would have been observed had the subject been exposed to treatment $\overline{a}_k$ until time $k$ and treatment $0$ thereafter.  
	Implicit in this notation is the assumption of no interference between  different subjects.

	Following \cite{robins1986new}, we make  the sequential ignorability assumption such that 
	\begin{equation}
	\label{eqn:seq_igno_2time}A_k \;\ind\; Y(\overline{a}_{K}) \mid \overline{L}_k, \overline{A}_{k-1} = \overline{a}_{k-1}
	\end{equation}	
	for all $\overline{a}_K \in \{0,1\}^K$. 
We also make the positivity assumption such that 
	\begin{equation}
	\label{eqn:positivity}
	P(A_k = a_k \,|\, \overline{L}_k = \overline{l}_k, \overline{A}_{k-1} = \overline{a}_{k-1}) \in (0,1), k=0,\ldots, K 
	\end{equation}
	as long as $P(\overline{L}_k = \overline{l}_k, \overline{A}_{k-1} = \overline{a}_{k-1})>0$.

SNMMs model the conditional causal contrasts \citep{robins1994correcting}:
	\begin{flalign*}
	B(\overline{l}_k, \overline{a}_{k-1};{\bm \alpha}) &\equiv
	g\{E(Y(\overline{a}_{k-1}, 1,\bm 0) \mid \overline{L}_k=\overline{l}_k, \overline{A}_{k-1}=\overline{a}_{k-1}; \bm \alpha) \} 			 \\
	& \quad\quad\quad\quad\quad\quad   -\, g\{E(Y(\overline{a}_{k-1}, 0, \bm 0) \mid \overline{L}_k=\overline{l}_k, \overline{A}_{k-1}=\overline{a}_{k-1}; \bm \alpha) \} \numberthis \label{eqn:snmm}  
	\end{flalign*}
	for $k=0, \ldots, K$, where $B(\overline{l}_k, \overline{a}_{k-1};{\bm 0}) = 0.$
		The contrasts \eqref{eqn:snmm} are called \emph{causal blip functions} as they describe the effect of receiving a last `blip' of treatment at time $k$ and then not receiving treatment thereafter (versus not receiving treatment at times $k,\ldots , K$). 
		
	
	SNMMs are often used for analysis of dynamic treatment regimes, where a dynamic regime is one  in which a subject's
	treatment choice $A_k$ depends on the intermediate responses up to that point $\overline{L}_k$ and previous treatment history $\overline{A}_{k-1}$. In fact,  under sequentially ignorability \eqref{eqn:seq_igno_2time} and the positivity assumption \eqref{eqn:positivity},  the so-called g-null hypothesis
	\begin{equation}
	\label{eqn:g-null}
	\mathcal{H}_0: E[Y(g)] = E[Y] \quad \text{for all } g\in \mathbb{G},
	\end{equation}
	is equivalent to  all the contrasts in \eqref{eqn:snmm} being equal to 0,
	where $\mathbb{G}$ denotes the set of all generalized treatment regimes consisting of all non-dynamic and dynamic treatment regimes \citep{robins1994correcting}.
	This statement holds regardless of the specific modeling assumptions placed on \eqref{eqn:snmm}, since the SNMMs are guaranteed to be correctly specified under the g-null.
	
	\subsection{Structural nested mean models: Estimation}
	
    In practice, the SNMM parameters are often estimated using doubly robust g-estimation methods \citep{vansteelandt2014structural}; see Section S7.2 in the Supporting Information for a detailed discussion. Alternatively, they may be estimated using regression-based methods, in a way similar  to the GLM approach for estimating the SMM parameters. As an illustration, we discuss regression-based inference for the multiplicative SNMM with two time points and continuous outcome.
    
  	When $K=1$, the multiplicative SNMMs model the following causal blip functions:
    	\begin{flalign}
\text{(causal blips)} \quad	\theta_0(l_0) &\equiv \dfrac{E[Y(1,0)\mid L_0=l_0]}{E[Y(0,0)\mid L_0=l_0]}, \label{eqn:a0} \\
\theta_1(l_0,a_0,l_1) &\equiv \dfrac{E[Y(a_0,1)\mid L_0=l_0, A_0=a_0, L_1=l_1]}{E[Y(a_0,0)\mid L_0=l_0, A_0=a_0, L_1=l_1]} = \dfrac{E[Y\mid l_0, a_0, l_1, 1]}{E[Y\mid l_0, a_0, l_1, 0]}.  \numberthis \label{eqn:a1}
		\end{flalign}
To allow for regression modeling on $E[Y\mid l_0, a_0, l_1, a_1],$ one may 
specify additional nuisance models on the following functions \citep[Appendix 2, p.36]{robins1997causal}: 
	\begin{flalign}
		\text{($L_1$ blip)} \quad 	\widetilde{\phi}(l_0, a_0,l_1) &\equiv \dfrac{E[Y(a_0,0) \mid L_0=l_0, A_0 = a_0, L_1 = l_1]}{E[Y(a_0,0) \mid L_0=l_0, A_0 = a_0, L_1 = 0]} =  \dfrac{E[Y\mid l_0, a_0, l_1, 0]}{E[Y\mid l_0, a_0, 0, 0]};   \label{eqn:l1}\\
		\widetilde{\eta}(l_0, a_0) &\equiv E(L_1 \mid L_0 = l_0, A_0=a_0);   \label{eqn:l1-past}\\
		\widetilde{\psi}(l_0)& \equiv  E(Y\mid L_0 = l_0, A_0=0,L_1=0,A_1=0). \label{eqn:baseline2}
		\end{flalign}
	Note that the ``$L_1$ blip'' is not causal. Furthermore, with these nuisance models, the mean potential outcome $E[Y(a_0, a_1)]$ may be evaluated via the g-formula \eqref{eqn:g-formula2}.

	\section{Parameterizations of binary SNMMs}
	\label{sec:parameterization}
	
	
	We now consider the case with a binary outcome $Y$. In this case, regression models on $E[Y\mid l_0, a_0, l_1, a_1]$ and $f(L_1\mid l_0, a_0)$ give rise to likelihood functions. Hence, in principle, one may use likelihood-based inference for inferring the SNMM parameters. However, similar to the case of SMMs discussed in Section \ref{sec:smm}, with a binary $Y$, the models for \eqref{eqn:a0} -- \eqref{eqn:baseline2} are variation dependent on each other, leading to undesirable consequences for interpretation, estimation and computation. These problems may be avoided if (I) the SNMMs are variation independent of each other; (II) the nuisance models  are variation independent of the SNMMs; (III) for each value of $l_0$ there exists a bijection between the the  $(8+2)$ parameters $E[Y\mid l_0,a_0,l_1,a_1]$, $E(L_1\mid l_0,a_0)$ and the combination of SNMMs and nuisance models.   In Section \ref{subsec:var_ind_snmm}, we show that (I) is true for multiplicative SNMMs but not for additive SNMMs. As a result, in general, estimators for the additive SNMM parameters may not be obtained via unconstrained maximum likelihood estimation.  On the other hand, to construct an unconstrained MLE for the multiplicative SNMM parameters, in Section \ref{subsec:2time} and \ref{subsec:general}, we propose novel nuisance functions that satisfy criteria (II) and (III).
	
	\subsection{Variation independence of SNMM parameters}
	\label{subsec:var_ind_snmm}

	\begin{prop}
		\label{prop:var_ind}
		If $K\geq 1$, then	the additive SNMMs  are \emph{variation dependent} on each other, while the multiplicative SNMMs  are {variation independent} of each other.
	\end{prop}
	
	Proposition \ref{prop:var_ind} may be surprising at first sight. 
	To provide heuristics, we illustrate the result in the case                                                                             $K=1$. A formal proof will become obvious later given Theorem  \ref{thm:main_general}.
	
	When $K=1$,  the additive SNMMs model a sequence of contrasts including
	\begin{flalign*}
	& E[Y(1,0) - Y(0,0)\mid L_0 = l_0] \in (-1,1), \numberthis \label{eqn:additive_snmm1}  \\
	& E[Y(1,1)  -  Y(1,0) \mid  L_0 = l_0, A_0=1, L_1 = l_1]\in (-1,1).  \numberthis \label{eqn:additive_snmm2}
	\end{flalign*}
	Marginalizing \eqref{eqn:additive_snmm2} over the distribution of $L_1$ conditional on $L_0 = l_0, A_0=1$ and using the sequential ignorability assumption \eqref{eqn:seq_igno_2time}, we get
	\begin{equation}
	\label{eqn:additive_snmm3}
	E[Y(1,1)  -  Y(1,0)\mid L_0 = l_0]\in (-1,1).
	\end{equation}
	If \eqref{eqn:additive_snmm1} were variation independent of \eqref{eqn:additive_snmm2} (and hence \eqref{eqn:additive_snmm3}), then the range of the sum of \eqref{eqn:additive_snmm1} and \eqref{eqn:additive_snmm3} would be $(-2,2).$ This contradicts the fact that the range of		$E[Y(1,1) - Y(0,0)\mid L_0 = l_0]$ is $(-1,1).$
	
	The reasoning above does not constitute a contradiction for the multiplicative  SNMM as it  specifies a sequence of differences of the form $\log\{ E[Y(\overline{a}_k, 0)\mid \overline{L}_k=\overline{l}_k, \overline{A}_k=\overline{a}_k]\} - \log\{E[Y(\overline{a}_{k-1}, 0)\mid \overline{L}_k=\overline{l}_k, \overline{A}_k=\overline{a}_k]\} \in \mathbb{R}.$ 	
	In simple terms, the multiplicative SNMMs are variation independent as $\mathbb{R} + \mathbb{R} = \mathbb{R}$, whereas the additive SNMMs are variation {\em dependent} as $(-1,1) + (-1,1) \not\subset (-1,1)$;  here  interval additions are defined as $(x_1, x_2) + (y_1,y_2) = \{x+y\mid x\in (x_1, x_2), y\in (y_1,y_2) \} = (x_1+y_1, x_2+y_2).$ We provide further graphical illustrations in Section \ref{sec:geometric-interpretation} in the Supporting Information.

	\subsection{Parameterization for the multiplicative SNMM: The two time points case}
	\label{subsec:2time}
	
	We now discuss the choice of nuisance models for the binary multiplicative SNMM with two time points. We first note that given $\widetilde{\eta}(l_0, a_0)$, the SNMM parameters $\theta_0(l_0),  \theta_1(l_0,a_0,l_1)$ and $L_1$ blip function 
	$\widetilde{\phi}(l_0,a_0,l_1)$ imply the seven conditional relative risks $E(Y\mid l_0,a_0,l_1,a_1) / E(Y\mid l_0,0,0,0), (a_0,l_1,a_1)\in \{0,1\}^3 \setminus \{(0,0,0)\}$; see eqn. \eqref{eqn:theta0} in the Supporting Information.  As we discussed in Section \ref{sec:smm}, these conditional relative risks are variation dependent on the baseline risk $\widetilde{\psi}(l_0).$ 
	
	To solve this problem, motivated by  the $l_0$-specific odds product in \cite{richardson2017modeling}, we propose to replace the baseline risk function $\widetilde{\psi}(l_0)$ in \eqref{eqn:baseline2} with  the $l_0$-specific generalized odds product:
	\begin{equation*}
	    	\text{gop}(l_0) \equiv \dfrac{\prod\limits_{a_0=0,1}\prod\limits_{l_1=0,1}\prod\limits_{a_1=0,1} E[Y\mid l_0, a_0, l_1, a_1] }{\prod\limits_{a_0=0,1}\prod\limits_{l_1=0,1}\prod\limits_{a_1=0,1} (1-E[Y\mid l_0, a_0, l_1, a_1])}.
	\end{equation*}
	Note that similar to the $l_0$-specific odds product, the $l_0$-specific generalized odds product is only a function of $l_0$ and not $(a_0, l_1, a_1)$.
	Theorem \ref{thm:main} shows that replacing $\widetilde{\psi}(l_0)$ with $\text{gop}(l_0)$ gives rise to a variation independent parameterization.

	\begin{theorem}
		\label{thm:main}
		Suppose that the sequential ignorability assumption \eqref{eqn:seq_igno_2time} holds. 	Let $\mathcal{M}$ denote the  model specified by the SNMMs \eqref{eqn:a0} and \eqref{eqn:a1}, a model on $\text{gop}(l_0),$
		and models on the following nuisance parameters:
		\begin{flalign*}
	\text{($L_1$ {\rm blip})} \quad 	\phi(l_0, a_0) &\equiv \widetilde{\phi}(l_0, a_0, 1), a_0 = 0,1, \numberthis \label{eqn:l1-blip-thm1}\\
			\eta(l_0, a_0) &\equiv E[L_1 \mid L_0 = l_0, A_0=a_0], a_0 = 0,1.
		\end{flalign*}
Then for each $l_0,$ the map given by 
		\begin{flalign*}
		&	(\theta_0(l_0),\theta_1(l_0,1,1), \theta_1(l_0,1,0), \theta_1(l_0,0,1),\theta_1(l_0,0,0),\phi(l_0,0),\phi(l_0,1),\text{\rm gop}(l_0),\eta(l_0,0),\eta(l_0,1)) \rightarrow \\
		&\quad \quad \quad \quad	 (P(Y=1\mid l_0, a_0, l_1, a_1), a_0, l_1, a_1 \in \{0,1\}; \eta(l_0, 0), \eta(l_0, 1))
		\numberthis	\label{map:one-to-one}
		\end{flalign*}				
		is a bijection from $(\mathbb{R}^{+})^8 \times (0,1)^2$ to $(0,1)^{10}$.
		Furthermore, models defining $\mathcal{M}$ are variation independent of each other. 
	\end{theorem}

	Theorem \ref{thm:main} may be generalized to accommodate categorical or (under additional monotonicity conditions) continuous time-varying treatments $A_0, A_1$ and confounder $L_1$; see Section \ref{sec:extension-general-treatment} in the Supporting Information for details.

	\subsection{Parameterization for the multiplicative SNMM: The general case}
	\label{subsec:general}
	
	To describe parameterizations for the binary multiplicative SNMM in the general case, we first introduce some notation: \begin{flalign*}
	\vec{0}_k  &\equiv \overbrace{(0, \ldots, 0)}^{k \ 0s}; \\
	\hbox{for } k=0,\ldots ,K:\quad E[Y(\overline{a}_k, \bm 0)] &\equiv E[Y(\overline{a}_k, \vec{0}_{K-k})]; \\
	\hbox{for } k=0,\ldots ,K:\quad \theta_k(\overline{a}_{k-1}, \overline{l}_k)  &\equiv	{\dfrac{E[Y(\overline{a}_{k-1},1, \bm 0) \mid \overline{A}_{k-1}=\overline{a}_{k-1}, \overline{L}_k=\overline{l}_k]}{E[Y(\overline{a}_{k-1}, 0, \bm 0) \mid \overline{A}_{k-1}=\overline{a}_{k-1}, \overline{L}_k=\overline{l}_k]};}  \\
	\hbox{for } k=0,\ldots ,K-1:\quad\phi(\overline{a}_k, \overline{l}_k) &\equiv \dfrac{E[Y(\overline{a}_k,\bm  0)\mid \overline{A}_k=\overline{a}_k, \overline{L}_k=\overline{l}_k, L_{k+1} = 1]}{E[Y(\overline{a}_k,\bm 0)\mid \overline{A}_k=\overline{a}_k, \overline{L}_k=\overline{l}_k, L_{k+1} = 0]}; \\
\hbox{for } k=0,\ldots ,K-1:\quad \eta(\overline{a}_k, \overline{l}_k) &\equiv E[L_{k+1}\mid \overline{A}_k = \overline{a}_k, \overline{L}_k = \overline{l}_k]. 
	\end{flalign*}
	We also let $\overline{a}_{-1} = \emptyset$ so that when $k=0$,
	$$
	\theta_k(\overline{a}_{-1}, \overline{l}_0)  \equiv	\dfrac{E[Y(1, \bm 0) \mid  {L}_0={l}_0]}{E[Y(0, \bm 0) \mid  {L}_0={l}_0]} \quad \hbox{ and }
	\phi(\overline{a}_0, \overline{l}_0) \equiv \dfrac{E[Y(a_0,\bm  0)\mid A_0=a_0, {L}_0={l}_0, L_{1} = 1]}{E[Y(a_0,\bm 0)\mid A_0=a_0, {L}_0={l}_0, L_{1} = 0]}.
	$$
	The following Theorem \ref{thm:main_general} gives the general form of our nuisance parameters for binary multiplicative SNMMs.
	
	\begin{theorem}
		\label{thm:main_general}
		Suppose that the sequential ignorability assumption \eqref{eqn:seq_igno_2time} holds. 	Let $\mathcal{M}$ denote the  model specified by the SNMMs on
		\begin{flalign*}
		\text{(Stage-k causal blip)} \quad	\bm{\theta} = 	(\theta_k(\overline{a}_{k-1}, \overline{l}_k): k=0,\ldots, K)
		\end{flalign*}
		and models on the following nuisance parameters
		\begin{flalign*}
			\text{($L_{k+1}$ blip)} \quad\bm{\phi} &= (\phi(\overline{a}_k, \overline{l}_k): k=0,\ldots,K-1); \\
		\bm{\eta} &= (\eta(\overline{a}_k, \overline{l}_k): k=0,\ldots,K-1); \\
		 GOP(l_0) &\equiv \dfrac{\prod\limits_{\overline{a}_K,{l}_1,\ldots, l_K} E[Y\mid \overline{A}_K=\overline{a}_K, \overline{L}_K=\overline{l}_K] }{\prod\limits_{\overline{a}_K,{l}_1,\ldots ,l_K} (1-E[Y\mid \overline{A}_K=\overline{a}_K, \overline{L}_K=\overline{l}_K])}.
		\end{flalign*}
	Then for each $l_0$, the map given by 
		\begin{equation}
		\label{eqn:mapping_bivariate}
		(\bm{\theta},\bm{\phi},GOP,\bm{\eta}) \rightarrow (E[Y\mid  \overline{A}_K=\overline{a}_K, \overline{L}_K=\overline{l}_K], \bm{\eta})
		\end{equation}				
		is a bijection from $(\mathbb{R}^{+})^{d_1} \times (0,1)^{d_2}$ to $(0,1)^{d}$,
		where $d_1 = 2^{2K+1}, d_2 = \sum\limits_{k=0}^{K-1} 2^{2k+1}, d=d_1+d_2$.
		 Furthermore, models in $\mathcal{M}$ are variation independent of each other.
	\end{theorem}

	\section{Estimation}
	\label{sec:estimation}
	\subsection{Maximum likelihood estimation: The two time points case}
	\label{sec:mle}
	
	We first discuss maximum likelihood estimation for a multiplicative SNMM with two time points. 
	Suppose models in $\mathcal{M}$ are specified up to a finite dimensional parameter, then  the parameters may be estimated directly via unconstrained maximum likelihood based on  the bijection \eqref{map:one-to-one}. For example, in the simulations, we assume that 
	\begin{flalign}
	\theta_0(l_0) &= \exp(\alpha_{j}^T l_0), j=0;\label{eqn:simu3} \\
	\theta_1(l_0, j) &= \exp(\alpha_j^T l_0), j=(1,1),(1,0),(0,1),(0,0); \label{eqn:simu5}\\
	\phi(l_0, j) &= \exp(\beta_{j}^T l_0), j=0,1;\label{eqn:simu6} \\
	\text{gop}(l_0) &= \exp(\delta^T l_0),\label{eqn:simu7} \\
	\eta(l_0, j) &= \text{expit} (\gamma_j^T l_0), j = 0,1;\label{eqn:simu2} 
	\end{flalign}
 For every possible value of $\alpha_j, \beta_j, \delta, \gamma_j,$ we may use equations  \eqref{eqn:simu3} -- \eqref{eqn:simu2} and the following Algorithm  \ref{alg:estimate_prob_k1} to compute $E(Y\mid l_0,a_0,l_1,a_1)$. The likelihood is then calculated as $\prod\limits_{i=1}^n L_{i}^Y \times L_{i}^{L}$, where 
	$
	    L_{i}^Y = E(Y\mid L_{0i},A_{0i},L_{1i},A_{1i})^{Y_i}\left\{ 1-E(Y\mid L_{0i},A_{0i},L_{1i},A_{1i}) \right\}^{1-Y_i}
	$
	and
	$
	    L_{i}^L = E(L_1\mid L_{0i}, A_{0i})^{L_{1i}} \left\{ 1-E(L_1\mid L_{0i}, A_{0i}) \right\}^{1-L_{1i}}.
	$
	The unconstrained maximum likelihood estimate of $(\alpha_j,$ $\beta_j, \delta, \gamma_j)$ can then be obtained by directly maximizing the log likelihood.

		\begin{algorithm}
	\caption{\ \ Compute $E[Y\mid l_0,  a_0, l_1,a_1]$ from $(\theta_0(l_0), \theta_1(l_0,a_0,l_1), \phi(l_0,a_0), gop(l_0), \eta(l_0,a_0))$}
		\label{alg:estimate_prob_k1}
		\begin{enumerate}
		\item Compute $\dfrac{E[Y\mid l_0, a_0, l_1, 1]}{E[Y\mid l_0, a_0, l_1, 0]}$, $\dfrac{E[Y\mid l_0, a_0, 1, 0]}{E[Y\mid l_0, a_0, 0, 0]}$ via 
		\eqref{eqn:a1} and \eqref{eqn:l1}, and then compute $\dfrac{E[Y\mid l_0, 1, 0, 0]}{E[Y\mid l_0, 0, 0, 0]}$ as \begin{flalign*}
\dfrac{E[Y\mid l_0, 1, 0, 0]}{E[Y\mid l_0, 0, 0, 0]} &=   \theta_0(l_0)  \dfrac{\eta(l_0,0)\phi(l_0,0) + 1-\eta(l_0,0)}{\eta(l_0,1)\phi(l_0,1) + 1-\eta(l_0,1)},
			\end{flalign*}
which holds by \eqref{eqn:a0} and \eqref{eqn:l1-blip-thm1} since
\begin{flalign*}
			    \theta_0(l_0) &= \dfrac{\eta(l_0,1)E[Y\mid l_0, 1,1,0] + (1-\eta(l_0,1)) E[Y\mid l_0, 1,0,0]}{\eta(l_0,0)E[Y\mid l_0, 0,1,0] + (1-\eta(l_0,0)) E[Y\mid l_0, 0,0,0]} \\
			    &= \dfrac{\eta(l_0,1)\phi(l_0,1) + 1-\eta(l_0,1)}{\eta(l_0,0)\phi(l_0,0) + 1-\eta(l_0,0)} \times \dfrac{E[Y\mid l_0, 1, 0, 0]}{E[Y\mid l_0, 0, 0, 0]}.
			\end{flalign*}
			\item Compute $r_{a_0,l_1,a_1}(l_0) \equiv \dfrac{E[Y\mid l_0, a_0, l_1, a_1]}{E[Y\mid l_0,0,0,0]}$ sequentially by \\[5pt]
			\begin{flalign*}
			\footnotesize \dfrac{E[Y\mid l_0, a_0, 0, 0]}{E[Y\mid l_0, 0, 0, 0]} &  \quad \text{\normalsize obtained from step (1)}; \\
			\footnotesize \dfrac{E[Y\mid l_0, a_0, l_1, 0]}{E[Y\mid l_0, 0, 0, 0]} &=  \footnotesize  \dfrac{E[Y\mid l_0, a_0, 1, 0]}{E[Y\mid l_0, a_0, 0, 0]} \times \dfrac{E[Y\mid l_0, a_0, 0, 0]}{E[Y\mid l_0, 0, 0, 0]}; \\
				\footnotesize \dfrac{E[Y\mid l_0, a_0, l_1, a_1]}{E[Y\mid l_0, 0, 0, 0]} &=  \footnotesize  \dfrac{E[Y\mid l_0, a_0, l_1, a_1]}{E[Y\mid l_0, a_0, l_1, 0]} \times \dfrac{E[Y\mid l_0, a_0, l_1, 0]}{E[Y\mid l_0, 0, 0, 0]}. 
			\end{flalign*}\\[-4pt]
			\item Compute
			$
			r_{\text{max}}(l_0) = \max\limits_{a_0, l_1, a_1}  r_{a_0,l_1,a_1}(l_0).$\\[-4pt]
			\item Compute
			$
			k_{a_0, l_1, a_1}(l_0) \equiv 		\dfrac{	r_{a_0, l_1, a_1}(l_0)}{ r_{\text{max}}(l_0)}.
			$
			\\[-4pt]
			\item Let $p_{l_0, a_0, l_1, a_1} = E[Y\mid l_0, a_0, l_1, a_1]$. Suppressing dependence on $l_0,$ for $x\in (0,1),$ let
			$$
			g(x) = \sum\limits_{i=(a_0,l_1,a_1)} \log(k_i) + 8 \log(x)  - \sum\limits_{i=(a_0,l_1,a_1)}  \log(1-k_i x) - \log(gop).$$
			Find  the unique root of $g(x)$ in the interval $(0,1)$.  Set $p_{\text{max}}(l_0)$ to be this root.\\[-4pt]
			\item  Compute 
			$
			E[Y\mid l_0, a_0, l_1, a_1] = k_{a_0, l_1, a_1}(l_0) \times p_{\text{max}}(l_0).
			$
		\end{enumerate}
	\end{algorithm}
	
		Alternatively,  a two-step procedure may be employed, in which one first estimates $\gamma_j, j=0,1,$  by maximizing the likelihood associated with $P(L_1 = 1\mid A_0, L_0),$ that is, $\prod\limits_{i=1}^n L_{i}^L.$
The value of $\gamma_j, j=0,1$ are then taken as fixed before proceeding to find the partial maximum likelihood estimate of $\alpha_j, \beta_j, \delta$ that maximizes $\prod\limits_{i=1}^n L_{i}^Y.$
 Inference for both the unconstrained and two-step maximum likelihood estimates can then be performed based on the non-parametric bootstrap.

	\subsection{Maximum likelihood estimation: The general case}

	In the general case with $K+1$ time points, in principle, the unconstrained and two-step maximum likelihood estimate may be obtained in a similar way. 
	The corresponding algorithm to compute the mapping \eqref{eqn:mapping_bivariate} is given in 
	Algorithm \ref{alg:estimate_prob} in the Supporting Information.
	
	
	However, note that in general,  the dimension of model parameters $(\bm{\theta}, \bm \phi, GOP, \bm{\eta})$ grows exponentially with $K$. To avoid possible identification problems with large numbers of follow-ups and moderate sample sizes, in practice one may make further dimension reducing assumptions on $\mathcal{M}$. In the data application, we make the  Markov assumption that  $\bm{\theta}, \bm{\phi}$ and $\bm\eta$  depend on the past history only through the most recent $L_k, A_k$ and $(A_k, L_k)$, respectively,  and assume that such dependencies are homogeneous over time:
	\begin{equation}
	\label{eqn:markov}
	\theta_k(\overline{a}_{k-1}, \overline{l}_k) =  \theta({l}_k),\quad
	\phi(\overline{a}_k, \overline{l}_k) = \phi({a}_k),\quad
	\eta(\overline{a}_k, \overline{l}_k) = \eta(a_k, l_k).	\end{equation}
	Similar assumptions have also been invoked in recent work by \cite{qian2021estimating}. 
	The Markov assumption may be extended to allow dependence on the previous two time points; see Section \ref{sec:markov-extension} in the Supporting Information for details.
	
	Furthermore, since the dimension of $E[Y\mid \overline{A}_K, \overline{L}_K]$ grows exponentially with $K$, Algorithm \ref{alg:estimate_prob}, and specifically Steps 2 -- 5 may be computationally prohibitive even when $K$ is small to moderate. 
	To resolve this problem, instead of computing $r_{\overline{a}_K, \overline{l}_K}$ for each $(\overline{a}_K, \overline{l}_K),$  we develop a dynamic programming method to compute the exact value of $r_{\text{max}}(l_0)$; the details are provided in Section \ref{sec:r_max} of the Supporting Information. Dynamic programming is applicable here due to our Markov assumption that $\bm{\theta}, \bm{\phi}$ and $\bm\eta$  depend on the past history only through the most recent $L_k$ and/or $ A_k.$ Dynamic programming   has also been used before in  the optimal structural nested models \citep[a.k.a. A-learning]{robins2004optimal} literature.  Furthermore, in Step 5, we approximate $g(x)$ by 
	$
	h_m(x) = \dfrac{d_1}{m} \sum\limits_{i = 1}^m  \log(k_i)  + d_1 \log(x)  - \dfrac{d_1}{m} \sum\limits_{i=1}^m \log(1-k_ix) - \log(GOP),	
	$
	where $i = 1,\ldots, m$ are  random samples drawn from a uniform distribution on the set $\{0,1\}^{d_1}.$ 
	To choose $m$ in practice, one may start with a small number, say $m=100$, and then increase $m$ until $h_m(x)$ is stable up to a threshold specified a priori.
	With the dynamic programming method and Monte Carlo approximation, the computational cost is reduced from $\mathcal{O}(\exp(K))$ to $\mathcal{O}(K)$.



	\subsection{Doubly robust estimation}
	\label{sec:dr}
	
	The proposed parameterization in Theorems \ref{thm:main} and \ref{thm:main_general} may be used to construct truly doubly robust estimators that are asymptotically linear if either the nuisance models  $\bm \phi(\beta), GOP(\delta), \bm \eta(\gamma)$ or the propensity score models $P(A_k = 1\mid \overline{A}_{k-1}, \overline{L}_k; \epsilon)$ are correctly specified. These estimators are called ``truly'' doubly robust because, as shown in Theorem \ref{thm:main}, the nuisance models $\bm \phi(\beta), GOP(\delta), \bm \eta(\gamma)$ are compatible with  the causal models $\bm \theta(\alpha)$. 
	To keep the exposition simple, we only discuss the case with $K=1$ here; the results can be extended to the general case, as we illustrate in Section \ref{sec:dr-data} in the Supporting Information.
	
	Specifically, let 
    $U_1(\alpha)    =  Y \theta_1(L_0, A_0,L_1;\alpha)^{-A_1}, 
    U_0(\alpha) = Y \theta_1(L_0, A_0,L_1;\alpha)^{-A_1} \theta_0(L_0; \alpha)^{-A_0}.$
Also let $\widehat{\alpha},\widehat{\beta}, \widehat{\delta}, \widehat{\gamma}$ and $\widehat{\epsilon}$ be preliminary estimates of $\alpha, \beta, \delta, \gamma$ and $\epsilon$ obtained via MLE or 2-step MLE, and
	let $\widehat{\alpha}_{dr}$ solve the following estimating equation \citep{vansteelandt2014structural}:
	\begin{flalign*}
	&\mathbb{P}_n \left( \left[ d_0(L_0, A_0) - \widehat{E}\left\{d_0(L_0, A_0) \mid L_0 \right\} \right] \times \left[ U_0 (\alpha) - \widehat{E}\left\{ U_0(\alpha) \mid L_0 \right\}  \right] + \right.\\
&	\left. \left[ d_1(L_0, A_0, L_1, A_1) - \widehat{E}\left\{d_1(L_0, A_0, L_1, A_1) \mid L_0, A_0, L_1 \right\} \right] \times \left[ U_1 (\alpha) - \widehat{E}\left\{ U_1(\alpha) \mid L_0, A_0, L_1 \right\}  \right]\right) =0, 
\numberthis\label{eqn:dr-ee}
\end{flalign*}
where $\mathbb{P}_n$ denotes the empirical mean operator: $\mathbb{P}_n O = \sum\limits_{i=1}^n O_i / n,$ $d_0(L_0, A_0)$ and $d_1(L_0, A_0, L_1, A_1)$ are measurable functions of the same dimension as $\alpha,$ and
\begin{flalign*}
    \widehat{E}\left\{ U_1(\alpha) \mid L_0, A_0, L_1 \right\} &= \widehat{E}\{Y(A_0,0)\mid L_0, A_0, L_1\} =  E(Y\mid L_0, A_0, L_1, A_1=0; \widehat{\alpha}, \widehat{\beta}, \widehat{\delta}, \widehat{\gamma}); \\
    \widehat{E}\left\{ U_0(\alpha) \mid L_0 \right\} &= \widehat{E}\{Y(0,0)\mid L_0\} = \eta(0; \widehat{\gamma}) E(Y\mid L_0,  A_0=0, L_1=1, A_1=0; \widehat{\alpha}, \widehat{\beta}, \widehat{\delta}, \widehat{\gamma} )  \\
    &  +\,\{1-\eta(0; \widehat{\gamma})\} E(Y\mid L_0, A_0=0, L_1=0, A_1=0; \widehat{\alpha}, \widehat{\beta}, \widehat{\delta}, \widehat{\gamma} ). \numberthis\label{eqn:22}
\end{flalign*}
 Note that for $k=0,1,$ the terms $\widehat{E}\left\{d_k(\overline{L}_k, \overline{A}_k) \mid \overline{L}_k, \overline{A}_{k-1} \right\}$ depend on the propensity score estimates $P(A_k=1\mid \overline{L}_k, \overline{A}_{k-1}, \widehat{\epsilon}).$ 
	
\cite{robins1994correcting} showed that under correct models for $\bm\theta(\alpha)$ and additional regularity conditions,  $\widehat{\alpha}_{dr}$ is  consistent and asymptotically
normally distributed provided that either the models for  \\ $E\left\{  U_k(\alpha) \mid \overline{L}_k, \overline{A}_{k-1}\right\}$ or $P(A_k=1\mid \overline{L}_k, \overline{A}_{k-1})$ are correctly specified. 	The optimal choice of $d_0(L_0, A_0)$ and $d_1(L_0, A_0, L_1, A_1)$ are given in Section \ref{sec:dr-simu} of the Supporting Information.


	\section{Simulation studies}
	\label{sec:data}
	
	We now evaluate the finite sample performance of our estimators with synthetic data. In our simulations, we consider two choices for the baseline covariates $L_0$: ``binary $L_0$'' that includes an intercept and a binary random variable generated from a Bernoulli distribution with mean $1/2$, or ``continuous $L_0$'' that includes an intercept and a random draw from the uniform distribution on $[-2,2].$  Conditional on $L_0$, the treatments $A_0, A_1$ and intermediate covariate $L_1$ were generated from \eqref{eqn:simu2}  and the following models:
	\begin{flalign}
	P(A_0 = 1\mid L_0) &= \text{expit}(\epsilon_{1}^T L_0); \label{eqn:simu1}\\
	P(A_1 = 1\mid L_1, A_0, L_0) &= \text{expit}(\epsilon_{2}^T L_0 + \epsilon_{3} A_0 + \epsilon_4 L_1), \label{eqn:simu4} \end{flalign}
	where $\epsilon_1 = \epsilon_2 =  (0.1, -0.5)^T, \epsilon_3 = 0.1, \epsilon_4 = -0.5,
	\gamma_0 = \gamma_1 = (-0.5, 0.1)^T.$
	The outcome $Y$ was generated indirectly through  models \eqref{eqn:simu5} -- \eqref{eqn:simu7},
	where $\alpha_j = (0,0.7)^T$ for $j = 0, (0,0), (0,1)$, $(1,0), (1,1)$; $\beta_0 = \beta_1 =(-0.5,0.1)^T;  \delta = (-0.5,1)^T.$

	We compare three estimation methods: 1) MLE as described in Section \ref{sec:mle}, in which we use the {\tt R} function {\tt optim} to maximize the likelihood, and {\tt uniroot} to find the solution in Step (5) of Algorithm \ref{alg:estimate_prob_k1}; 2) 2-step MLE as described in Section \ref{sec:mle}; 3) DR estimation as described in Section \ref{sec:dr}. To solve the doubly robust estimating equation \eqref{eqn:dr-ee}, we use {\tt optim} to minimize the square of eqn. \eqref{eqn:dr-ee}, in which we use the 2-step MLE estimates as the preliminary estimates for $\alpha, \beta, \delta, \gamma$ and  the  starting value for $\alpha$. The weighting functions are chosen as the optimal weighting function detailed in Section \ref{sec:dr-simu} of the Supporting Information. The propensity score model parameters $\epsilon_j, j=1,\ldots, 4$ are estimated using logistic regressions.
	As a summary measure, we also report estimates of the causal contrast $E[Y(1,1)]/E[Y(0,0)]$ using the g-formula \eqref{eqn:g-formula2}, 
	where $E(Y\mid l_0,a_0,l_1,a_1)$ and $f(l_1\mid l_0, a_1)$ were estimated using the models on $\theta_0(l_0),\theta_1(l_0,a_0,l_1), \phi(l_0,a_0),$  $\text{gop}(l_0), \eta(l_0,a_0).$
	{The true value for $E[Y(1,1)]/E[Y(0,0)]$ is evaluated via a simulation sample of size 100,000.}
	Unless otherwise specified, the simulation results are based on 500 Monte-Carlo runs of $n=1000$ units.

	Table \ref{tab:est}  summarizes the simulation results for binary $L_0$. Results with continuous $L_0$ are deferred to Table \ref{tab:est-L0cont} in the Supporting Information.  All methods  yield estimators with small biases relative to their standard errors, confirming consistency of the proposed estimators. The estimates of the 2-step MLE are very close to those of the MLE,
	which suggests that the conditional distribution of the outcome $Y$, i.e. $P(Y=y \mid A_1, L_1, A_0, L_0)$ contains little information on $\gamma_0$ and $\gamma_1$ relative to $P(L_1=l_1\mid A_0, L_0).$ As expected, the variance of the doubly robust estimator is no smaller than that of the maximum likelihood estimators.

	\begin{table}
		\begin{center}
			\caption{Bias $\times$ 100 (Monte Carlo standard error $\times$ 100) of the  proposed methods with a binary baseline covariate $L_0$. 
				The sample size is  1000}
			\bigskip
			\label{tab:est}
		{	\begin{tabular}{rccccccccc}
				\toprule
				&       \multicolumn{2}{c}	{MLE} &  \multicolumn{2}{c}	{2-step MLE} &  \multicolumn{2}{c}	{DR}   \\
				\cmidrule(r){2-3} \cmidrule(l){4-5} \cmidrule(l){6-7} 
				& \multicolumn{1}{c}{baseline} & \multicolumn{1}{c}{slope} &   \multicolumn{1}{c}{baseline} & \multicolumn{1}{c}{slope} &   \multicolumn{1}{c}{baseline} & \multicolumn{1}{c}{slope}\\
				\midrule
				\multicolumn{2}{l}{SNMM parameters }	 & & & \\
				$\theta_0(l_0)$ & 0.63(0.55) & -1.8(0.97) & 0.63(0.55) & -1.8(0.98) & 0.49(0.55) & 1.4(0.97) \\ 
  $\theta_1(l_0,1,1)$ & -1.5(1.2) & 3.0(1.6) & -1.5(1.2) & 2.9(1.6) & -0.92(1.2) & 3.7(1.7) \\ 
  $\theta_1(l_0,1,0)$ & 0.80(0.57) & 0.60(0.74) & 0.79(0.57) & 0.68(0.75) & 0.87(0.57) & -1.0(0.74) \\ 
  $\theta_1(l_0,0,1)$ & 0.90(1.3) & 0.15(2.1) & 0.91(1.3) & 0.045(2.1) & 0.75(1.3) & 2.7(2.1) \\ 
  $\theta_1(l_0,0,0)$ & 0.53(0.60) & -0.40(1.1) & 0.53(0.60) & -0.45(1.1) & 0.40(0.60) & 1.7(1.1) \\ \\
  \multicolumn{2}{l}{Nuisance parameters}	 & & & \\
    $\phi_0(l_0)$ & -1.5(1.1) & -1.4(1.9) & -1.5(1.1) & -1.4(1.9) & $-$  & $-$\\ 
  $\phi_1(l_0)$ & -1.1(0.84) & -0.90(1.3) & -1.1(0.84) & -0.78(1.3) & $-$  & $-$ \\ 
  $\text{gop}(l_0)$ & -6.5(3.5) & 2.2(5.5) & -6.5(3.5) & 1.4(5.4) & $-$  & $-$\\ 
  $\eta_0(l_0)$ & -0.73(0.58) & 0.73(0.78) & -0.73(0.58) & 0.80(0.78) & $-$  & $-$ \\ 
  $\eta_1(l_0)$ & -0.63(0.56) & 0.43(0.87) & -0.63(0.56) & 0.31(0.87) & $-$  & $-$ \\ \\
  \multicolumn{2}{l}{Marginal causal parameters}	 & & & \\[2pt]
     $E[Y(0,0)]$ & -0.12(0.13) & $-$ & -0.12(0.13) & $-$ & -0.29(0.13) &  $-$ \\[3pt] 
  $E[Y(1,1)]$ & -0.18(0.14) & $-$ & -0.18(0.14) & $-$ & 0.15(0.14) &  $-$ \\[5pt] 
  {\small $\dfrac{E[Y(1,1)]}{E[Y(0,0)]}$} & 1.3(0.79) & $-$ & 1.3(0.79) & $-$ & 3.2(0.80) &  $-$ \\ 
				\bottomrule 
			\end{tabular}}
		\end{center}
	\end{table}
	
We also compare these three methods in terms of their computation time; the detailed results can be found in Table \ref{tab:time} in the Supporting Information. All the computation was done on a Lenovo SD650 NeXtScale server using an Intel 8268 ``Cascade Lake'' processor and 192GB RAM.  As expected, the computation time with continuous $L_0$ is longer as there are many more different combinations of covariate values in the sample.  





A reviewer suggested that with the doubly robust g-estimation procedure, when the propensity score models are correctly specified,  our proposed parametrization may lead to a more efficient estimator than a non-compatible one, even if the nuisance models are misspecified. Motivated by this, we consider an additional simulation setting with a (spurious) baseline covariate vector $\widetilde L_0$ that includes an intercept and an independent random draw from $Bern(1/2)$. 
We consider two mis-specifications for the baseline  models: (1) our proposed DR estimator, with the gop model misspecified as a function of $\widetilde{L}_0:$ $\text{gop}(L_0) = exp(\delta^T  \widetilde{L}_0)$; note that in our approach, the baseline function $E[Y\mid L_0, A_0, L_1, A_1]$ is estimated through the gop model, so it is also misspecified; (2) the DR estimator with logistic baseline models, where the terms $E(Y\mid L_0, A_0, L_1, A_1=0)$ in \eqref{eqn:22} are estimated using a logistic regression that assumes $\text{\logit} E[Y\mid L_0, A_0, L_1, A_1] =b_0 \widetilde L_0 + b_1 A_0 + b_2 L_1 + b_3 A_1L_1.$ 
Table \ref{tab:sim-logit-vs-mle} summarizes the results. The estimates from the DR estimator with a logistic baseline model  have larger finite sample bias and variability than the proposed DR estimator, especially those for  $\theta_1(l_0, 0,1).$ This provides some initial evidence for the benefits of our proposed DR estimator, in the case that the baseline models are misspecified.

\begin{table}
\begin{center}
\caption{Bias $\times$ 100 (Monte Carlo standard error $\times$ 100) of the proposed DR estimator, and the DR estimator with a logistic baseline model, under misspecification of baseline nuisance models. The propensity score models are correctly specified.  The sample size is  1000}
			\bigskip
			\label{tab:sim-logit-vs-mle}
\begin{tabular}{rccccccccc}
\toprule
				&       \multicolumn{2}{c}	{DR (Proposed)} &  \multicolumn{2}{c}	{DR (logistic baseline model)}   \\
				\cmidrule(r){2-3} \cmidrule(l){4-5} 
				& \multicolumn{1}{c}{baseline} & \multicolumn{1}{c}{slope} &   \multicolumn{1}{c}{baseline} & \multicolumn{1}{c}{slope} \\
				\midrule
				\multicolumn{1}{l}{SNMM parameters}	 & & & \\
$\theta_0(l_0)$ & 0.12(0.54) & 2.4(0.97) & 2.7(0.51) & -1.5(0.96) \\ 
  $\theta_1(l_0,1,1)$ & -0.26(1.2) & 2.5(1.7) & -3.0(1.2) & 6.9(1.7) \\ 
  $\theta_1(l_0,1,0)$ & 0.97(0.57) & -1.3(0.74) & 0.35(0.56) & -0.48(0.74) \\ 
  $\theta_1(l_0,0,1)$ & -0.28(1.3) & 5.3(2.1) & 24(5.9) & -21(6.7) \\ 
  $\theta_1(l_0,0,0)$ & 0.26(0.60) & 2.1(1.1) & 1.2(0.61) & 1.1(1.2) \\ 
   \bottomrule
\end{tabular}
\end{center}
\end{table}

In Section \ref{sec:g-null} of the Supporting Information, we present an alternative simulation study that compares our proposed two-step MLE and doubly robust estimators, versus the g-computation method under the g-null hypothesis that $\theta_0(l_0)  = \theta_1(l_0,a_0,l_1) = 1$ for all $l_0$ and $(a_0,l_1) = (1,1), (1,0), (0,1), (0,0)$. 

	\section{Application to the Mothers' Stress--Children's Morbidity study}
	\label{sec:real_data}
	
\begin{figure}
		\centering
		\includegraphics[width=.7\textwidth]{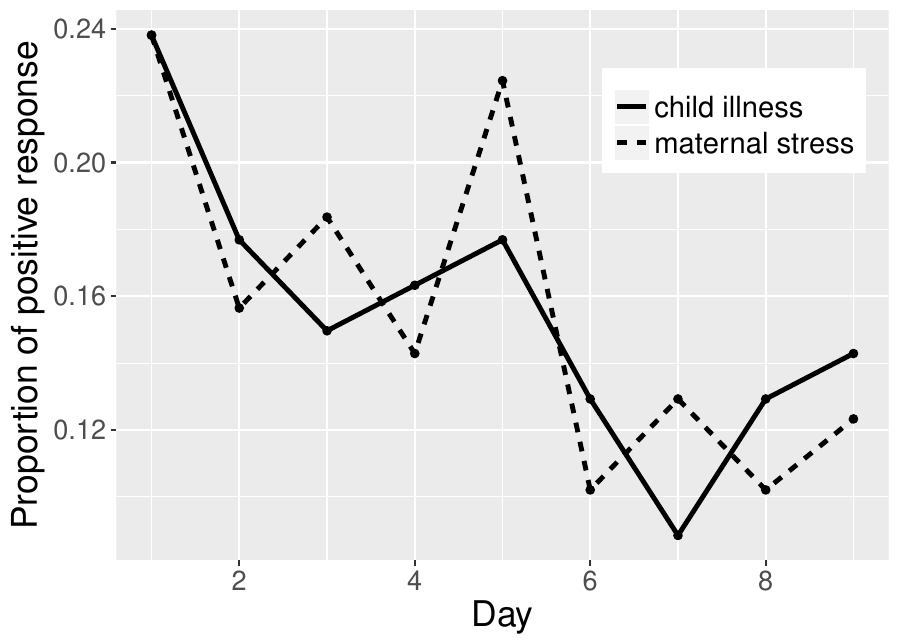}
		\caption{Mothers' stress and children's illness evolving over time.}
		\label{fig:descriptive}
	\end{figure}

	To illustrate the proposed methods, we reanalyze data from the Mothers' Stress-Children's Morbidity (MSCM) study \citep{zeger1986longitudinal},
	which consist of observations on 167 mothers with infants aged between 18 months and 5 years. Daily observations were taken on mothers' stress level and whether or not their child was ill. The total length of follow-up is 30 days.    
	Similar to \cite{robins1999estimation}, we are interested in whether or not maternal stress has an influence on child illness. Maternal stress may be considered as a treatment variable because it is a natural target for an intervention through support programs such as by providing  midwives. Following \cite{zeger1986longitudinal}, we use the first 9 days of records to illustrate use of the SNMM so that $K = 7$. The treatment variables are maternal stress indicators in the first 8 days, denoted as $A_0, \ldots, A_7$; the outcome of interest is whether or not the child is ill at the 9th day; the time-varying confounders are child illness in the first 8 days: $L_0, \ldots, L_7,$ and the time-independent baseline confounders include household size, employment, marital status and child's race. To distinguish 
	the time-independent confounders from the time-varying confounder measured at baseline, with slight abuse of notation, we use $X$ to denote the former, 
	and $L_0$ for the latter. 
	Note that the outcomes, time-varying confounders and treatments are all binary. In the data set made available to us, the time-independent confounders are also binary.  There are $147$ mother-child pairs with complete observations on all these variables and for illustrative purposes, we restrict our analysis to these pairs.
	Figure  \ref{fig:descriptive} shows the observations in the first 9 days. The correlation between maternal stress and children's illness, however,  may be subject to confounding by children's illness at earlier time points.  Our  formal causal analysis assumes that all confounders are measured, and that our parametric model specifications are correct.
	
	We assume the following causal blip models:
	\begin{flalign*}
	\log \dfrac{E[Y(\overline{a}_{k-1},1, \bm 0) \mid \overline{A}_k = \overline{a}_k, \overline{L}_k = \overline{l}_k, X]}{ E[Y(\overline{a}_{k-1}, 0, \bm 0) \mid \overline{A}_k = \overline{a}_k, \overline{L}_k = \overline{l}_k, X]} &= \alpha_0(1-l_k) + \alpha_1 l_k + \alpha_X^T X, k = 0,\ldots,K
	\end{flalign*}
	and the following nuisance models:
	\begin{flalign*}
	\log \dfrac{E[Y(\bm 0) \mid  L_{0}=1, X]} {E[Y(\bm 0) \mid  L_{0} = 0, X]} &= \beta_{L_0}  + \beta_X^T X, \\
	\log \dfrac{E[Y(\overline{A}_k, \bm 0) \mid \overline{A}_k, \overline{L}_k, L_{k+1}=1, X]} {E[Y(\overline{A}_{k}, \bm 0) \mid \overline{A}_k, \overline{L}_k, L_{k+1} = 0, X]} &= \beta_0(1-A_k) + \beta_1 A_k  + \beta_X^T X, \quad k=0, \ldots, K-1; \\
	GOP(L_0, X) &=  \delta_X^T X;\\
	\text{logit} (E[L_{k+1}\mid \overline{A}_{k},\overline{L}_k, X]) &= \gamma_{00}1(A_k=L_k=0) + \gamma_{01}1(A_k=0, L_k=1) +  \\
	\gamma_{10}1(A_k=1,&L_k=0) + \gamma_{11}1(A_k=L_k=1) + \gamma_X X,\; k=0,  \ldots, K-1; \\
	\text{logit} P(A_{k}=1\mid  \overline{L}_{k}, \overline{A}_{k-1}, X) &= \epsilon_0 + \epsilon_1 L_k + \epsilon_X X,\; k=1,  \ldots, K.
	\end{flalign*}
We apply three different estimation methods: 1) 2-step MLE as described in Section \ref{sec:mle}; (2)  DR estimation as described in Section \ref{sec:dr}, in which the weighting function is chosen as $d_m(\overline L_m, \overline A_m)  = A_m (1, L_m, X)^T$; (3) DR estimation with a logistic baseline model: 
\begin{flalign*}
\text{logit} (E[Y \mid \overline{A}_{K-1} = \overline{a}_{k-1}, A_K = 0, \overline{L}_K = \overline{l}_K, X]) &= \zeta_{00}1(a_{K-1}=l_K=0) + \zeta_{01}1(a_{K-1}=0, l_K=1) +  \\
\zeta_{10}1(a_{K-1}=1,&l_K=0) + \zeta_{11}1(a_{K-1}=l_K=1) + \zeta_X X.
\end{flalign*}
	Inference is based on 500 non-parametric bootstrap samples.  Table \ref{tab:DR-on-data} presents the analysis results.

	We first tested the g-null mean hypothesis \eqref{eqn:g-null}.
	Note that as $Y$ is binary, the g-null mean hypothesis coincides with the g-null hypothesis of \citet[\S 6]{robins1986new}.
	A valid level-$0.05$ test may be obtained by testing $(\alpha_0, \alpha_1, \alpha_X^T) = 0.$  
The p-values from the three methods all suggest  that we have  failed to reject the g-null (mean) hypothesis at the 0.05 level.
	Note that the standard generalized estimating equation approach of \citet{zeger1986longitudinal} cannot be used to test the g-null hypothesis in the presence of time-dependent confounding by $L_k$ \citep[\S 6]{robins1999estimation}.
	
	We then compare estimates of the causal parameters $(\alpha_0, \alpha_1, \alpha_X)$ from the three methods. The confidence interval of the proposed DR estimator is shorter than the DR estimator with a logistic baseline model. This is consistent with the findings in Table \ref{tab:sim-logit-vs-mle}, suggesting that our compatible parameterization may lead to efficiency gain in doubly robust estimation.  Consistent with the findings in Tables \ref{tab:est} and \ref{tab:est-L0cont} in the Supporting Information, the MLE leads to shorter confidence intervals than the DR estimators.

	
	We also compare the regime where all mothers are subject to substantial stress for 8 consecutive days, versus the regime where all mothers are never stressed during these 8 days, possibly due to a fully effective intervention program. We choose this comparison because it is expected to show the largest effect. We use estimates from the MLE as they are the most stable among the three. {Compared to the latter, the former regime is estimated to result in a 4.850 (95\% CI [1.202, 8.498]) fold  increase in risk of childhood illness on the 9th day, suggesting a statistically significant causal effect by comparing these extreme regimes.} 
	Note that, even though we failed to reject the overall g-null in a test with six degrees of freedom, we find statistical significance when comparing this particular pair, for which we anticipate the causal effect to be the largest.

		\begin{table}
		\begin{center}
			\caption{Estimation results for the MSCM study}
			\bigskip
			\label{tab:DR-on-data}
			\begin{tabular}{rccc}
				\toprule
				& MLE & DR (Proposed) & DR (logistic baseline model) \\ 
				\midrule
				p-value for testing  the g-null   &  0.793 & 0.213 & 0.692 \\ [10pt] 
		 \multicolumn{2}{l}{SNMM coefficients estimates with 95\% CI} && \\[5pt]
				{$\alpha_0$} & -0.15 (-0.98,0.67) & 2.01 (-5.91,9.93) & -1.35 (-10.09,7.39) \\ 
  {$\alpha_1$} & -0.28 (-1.08,0.53) & 1.77 (-5.63,9.18) & 0.67 (-8.59,9.93) \\[5pt]
  $\alpha_X$ &  && \\[2pt]
  Household size $>$ 3 & 0.30 (-0.40,0.99) & 1.06 (-5.74,7.86) & 1.52 (-6.11,9.15) \\ 
  Race non-white & 0.28 (-0.38,0.94) & 1.15 (-5.56,7.87) & 0.51 (-6.96,7.99) \\ 
  Employed & 0.08 (-0.41,0.58) & -4.35 (-10.95,2.26) & 0.29 (-7.97,8.56) \\ 
  Married & -0.15 (-0.61,0.30) & 2.47 (-8.58,3.63) & -0.47 (-7.91,6.98) \\[10pt] 
    Computation time* & 		24.95s & 234.49s & 205.4s \\
				\bottomrule  \\[10pt]
			\end{tabular}
		\end{center}
		
		*: Computation is done using the same resource as in the simulations, and the time reported is the time for getting the point estimate.
	\end{table}

	\section{Discussion}
	\label{sec:discussion}
	
	In this paper we introduce a general approach for causal inference from complex longitudinal data with binary outcomes and time-varying confounders. Our approach is based on the SNMMs developed by \cite{robins1994correcting}, which overcome the null paradox of the g-formula and have many important advantages over marginal structural models as detailed in Section \ref{sec:intro}. Furthermore, SNMMs provide natural non-centrality parameters to describe deviations from the causal g-null.
	When the outcome is unconstrained,  both the multiplicative SNMMs and additive SNMMs are variation independent of the conventional nuisance models as described in \citet[Appendix 2, p.36]{robins1997causal} and \cite{robins1994correcting}, respectively.
	However, the conventional multiplicative and additive SNMMs are not suitable for inferring causal effects with binary outcomes as they do not naturally respect the fact that probabilities are bounded above by one, whereas the logistic SNMMs cannot be used in combination with g-estimation methods that are guaranteed to yield a valid test of the g-null in randomized trials.  We address this problem in two ways: for binary multiplicative SNMMs we introduce novel nuisance models so that the SNMM parameters can be estimated in a compatible way; for binary additive SNMMs we show that the SNMMs are variation dependent on each other (and hence incompatible) so that additive SNMMs should probably be avoided when analyzing binary outcomes. 
	
	In this article, we have assumed that all the subjects have the same number of follow-ups. In practice, however, it is often the case that study participants receive different number of treatments due to loss to follow-up. {In this case, one may
	first create a pseudo-population by  reweighting the observed population with inverse probability of censoring weights, and then apply the proposed estimation methods to the pseudo-population; see \citet[][\S 21.5]{hernan2020causal} for a detailed discussion.}

	
	\section*{Acknowledgements}
	
	This research was supported in part by NSERC grants RGPIN-2019-07052, DGECR-2019-00453 and RGPAS-2019-00093, U.S. National Institutes of Health grants
	R01 AI032475, AI113251,  R01AA23187 and P41EB028242 and ONR grants N00014-15-1-2672 and N00014-19-1-2446. 
	The authors thank Robin Evans  for helpful comments, {and Susan Murphy for kind support.} The content is solely the responsibility of the authors and does not necessarily represent the official views of the funding agencies.

	  \section*{Data Availability Statement}
	  
The data that support the findings in this paper are openly available in Harvard Dataverse at \url{https://doi.org/10.7910/DVN/UZAI96} \citep{dataverse}.

	\thispagestyle{empty}
	\bibliographystyle{apalike}
	\bibliography{causal}

	\section*{Supporting Information}
	Web Appendices, Tables,  Figures and Proofs  referenced in Sections \ref{sec:intro} -- \ref{sec:real_data} are available with this paper at the Biometrics website on Wiley Online Library. The proposed CBS method is available both on Wiley Online Library and on Harvard Dataverse:  \cite{dataverse}.

\clearpage

 \centerline{\large\bf Supplementary Material for ``Coherent  modeling of }
\vspace{2pt}
 \centerline{\large\bf longitudinal causal effects on binary outcomes''}
\vspace{.25cm}
\vspace{.4cm} \centerline{Linbo Wang, Xiang Meng, Thomas S. Richardson, James M. Robins} \vspace{.4cm}  \vspace{.55cm} 
\par

\setcounter{equation}{0}
\setcounter{figure}{0}
\setcounter{table}{0}
\setcounter{section}{0}
\setcounter{algorithm}{0}
\renewcommand{\theequation}{S\arabic{equation}}
\renewcommand{\thefigure}{S\arabic{figure}}
\renewcommand{\thetable}{S\arabic{table}}
\renewcommand{\theassumption}{S\arabic{assumption}}
\def\thesection{S\arabic{section}}
\def\thealgorithm{S\arabic{algorithm}}

\begin{abstract}
Section \ref{sec:geometric-interpretation} provides a graphical illustrations of Proposition \ref{prop:var_ind}.
    Section \ref{sec:extension-general-treatment} contains extensions to general time-varying treatments and confounders. Sections \ref{appendix:proof} and \ref{section:proof2}  contain proofs of Theorems \ref{thm:main} and \ref{thm:main_general} in the main paper, respectively.  Section \ref{sec:algorithm} contains an algorithm for computing the mapping \eqref{eqn:mapping_bivariate} in the main paper. This is a generalization of Algorithm \ref{alg:estimate_prob_k1} in the main paper. Section \ref{sec:r_max} details the steps to compute $r_{\text{max}}$ in Step 3 of Algorithm \ref{alg:estimate_prob} using the dynamic programming algorithm  under a Markov assumption. Section \ref{sec:markov-extension} outlines a dynamic programming algorithm under a relaxed Markov assumption that allows for dependence on the previous two time points. Section \ref{sec:appendix-dr} provides details on the implementation of doubly robust estimation. Section \ref{sec:addition-simu} contains additional simulations.
\end{abstract}

\section{Graphical illustration of the odds product }

	\begin{figure}
		\begin{center}
			\begin{minipage}{0.45\textwidth}
				\begin{center}
					\includegraphics[width=\textwidth]{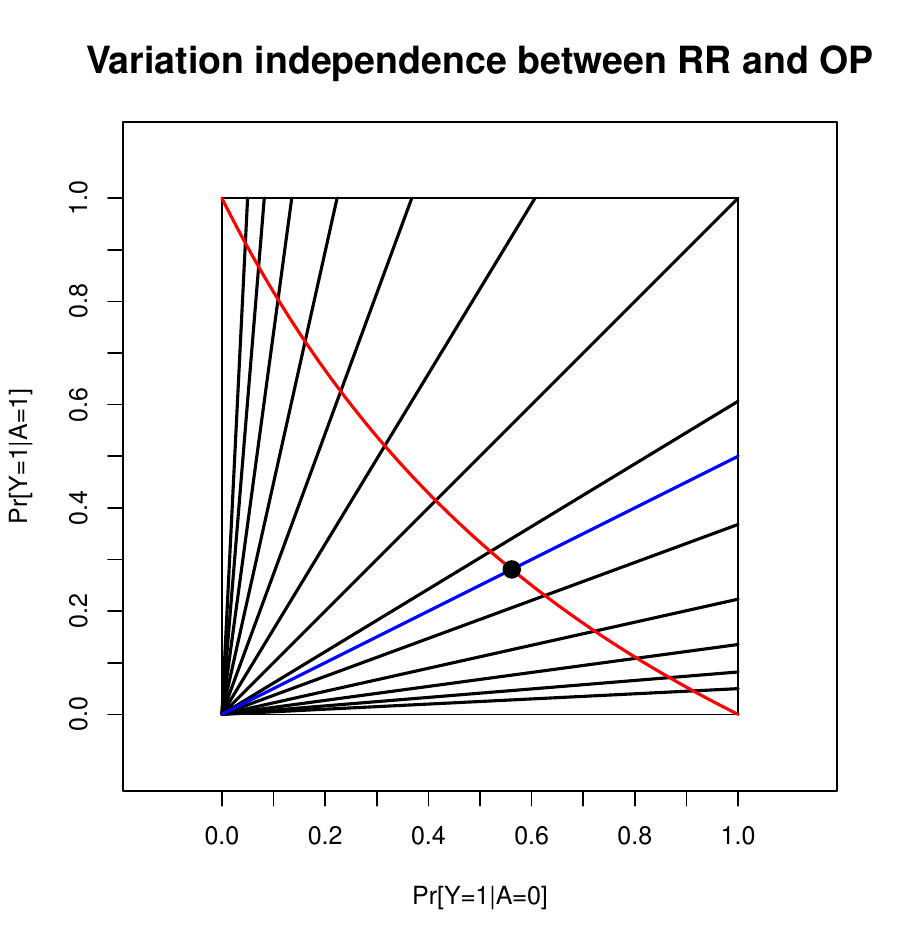}\\
				\end{center}
			\end{minipage}
			\begin{minipage}{0.45\textwidth}
				\begin{center}
			\includegraphics[width=\textwidth]{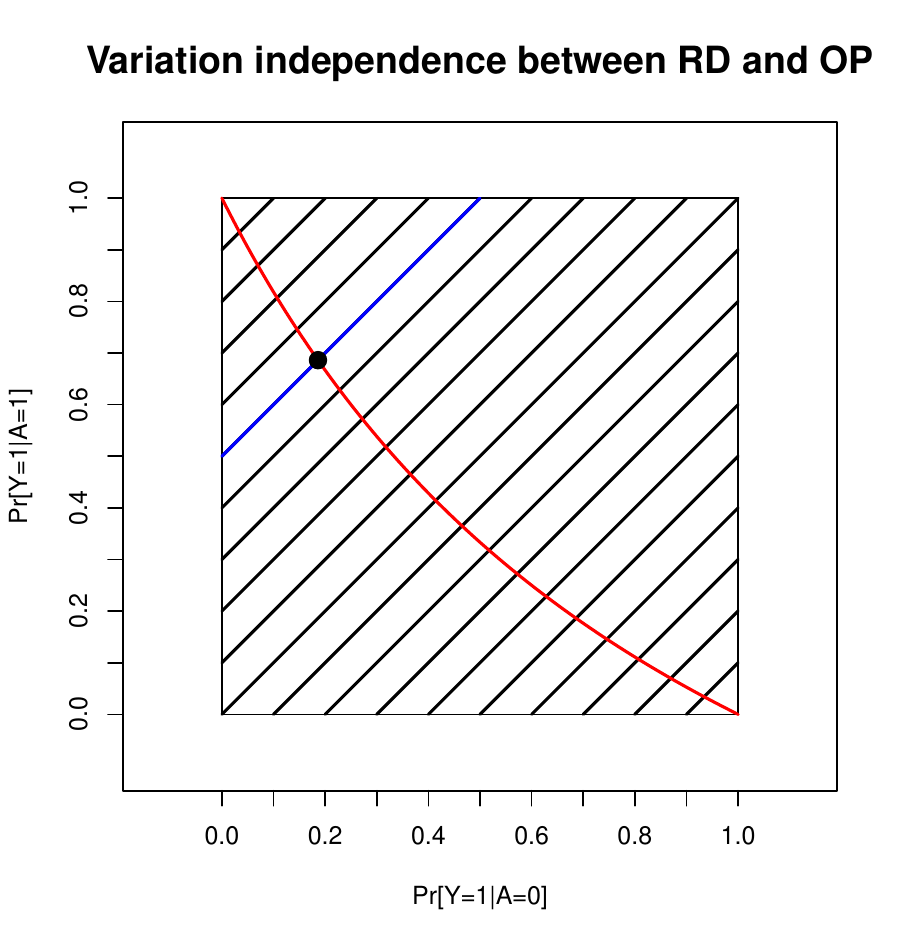}\\
				\end{center}
			\end{minipage}
				\begin{minipage}{0.45\textwidth}
				\begin{center}
			\includegraphics[width=\textwidth]{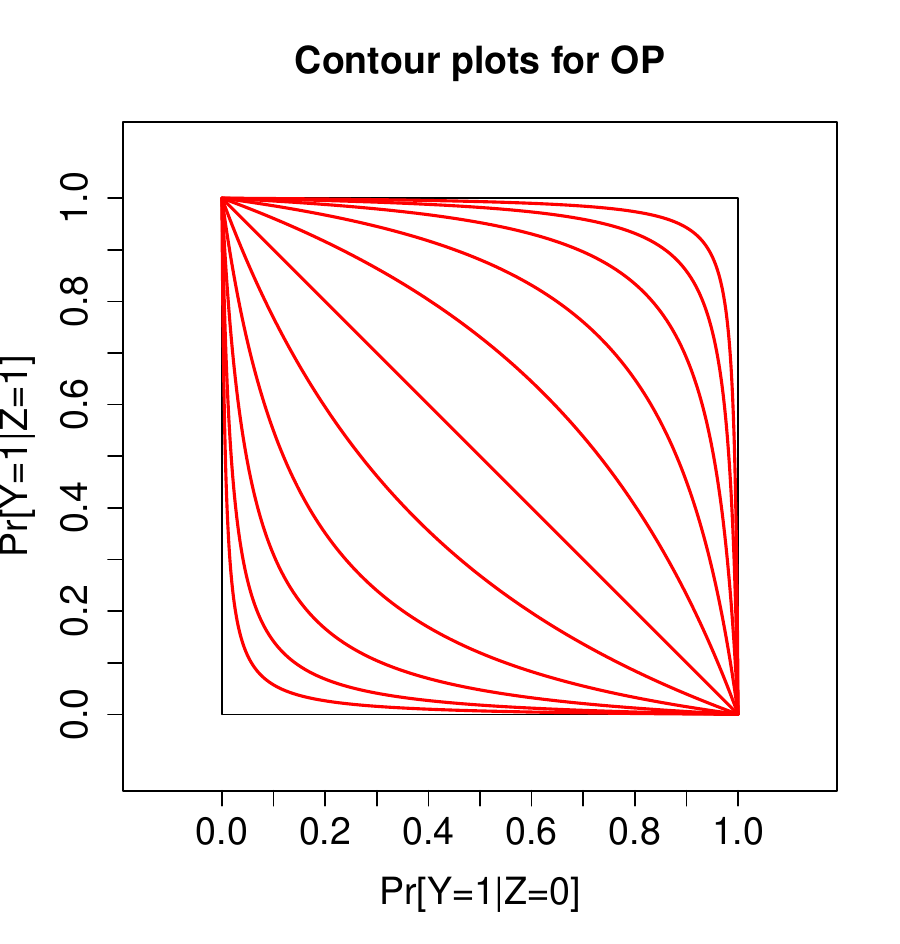}\\
				\end{center}
			\end{minipage}
		\end{center}
		\caption{L'Abb\'{e} plots: Lines of constant: (Upper left) log RR $\in (-3,-2.5,\ldots,  3)$, OP = 0.5 (red curve); (Upper right) RD $\in \{-0.9,-0.8,\ldots,0.9\}$,  OP = 0.5 (red curve); (Lower) log OP $\in \{-5,-4,\ldots,5\}$. 
			\label{fig:one}}
	\end{figure}

\section{Graphical illustrations of Proposition \ref{prop:var_ind}}
\label{sec:geometric-interpretation}

	We can   explain Proposition \ref{prop:var_ind} with the graphs in Figure \ref{fig:thomas-rd}.  
	As shown in Figure \ref{fig:thomas-rd} (a) and (c), with additive SNMMs, the second stage blips $E[Y(a_0,1)-Y(a_0,0)\mid L_0 = l_0, A_0=a_0, L_1=l_1], l_1=0,1$ may impose constraints on the second stage baseline quantities $E[Y(a_0,0)\mid L_0 = l_0, A_0=a_0, L_1=l_1]$. Marginalizing over the distribution of $L_1$ given $L_0, A_0$, these constraints may then imply constraints on $E[Y(a_0,0)\mid L_0 = l_0, A_0=a_0]$, which, by the sequential ignorability assumption, equals $E[Y(a_0,0)\mid L_0 = l_0]$. The constraints on $E[Y(a_0,0)\mid L_0 = l_0]$, $a_0=0,1$ are shown in the red and blue rectangles in Figure \ref{fig:thomas-rd} (b), respectively. The intersection of these rectangles, i.e. the dotted region in Figure \ref{fig:thomas-rd} (b),  defines the constraints on $(E[Y(0,0)\mid L_0 = l_0], E[Y(1,0)\mid L_0 = l_0])$ implied by the second stage blips.  One may see that some contour lines of $E[Y(1,0)-Y(0,0)\mid L_0 = l_0]$, such as the gray lines in Figure \ref{fig:thomas-rd} (b), have no intersection with the dotted feasible region for  $(E[Y(0,0)\mid L_0 = l_0], E[Y(1,0)\mid L_0 = l_0])$. This shows that the second stage blips may imply constraints on the possible values of the first stage blip $E[Y(1,0)-Y(0,0)\mid L_0 = l_0]$, so that they are variation dependent of each other. 
	
	We can apply the same reasoning to the multiplicative SNMMs, as illustrated in Figure \ref{fig:thomas-rr}. Given any values for the second stage blips, the feasible region for $(E[Y(0,0)\mid L_0 = l_0], E[Y(1,0)\mid L_0 = l_0])$ always includes the origin. Hence, unlike the case with additive SNMMs, this feasible region will always intersect with any contour lines for the first stage blip  $E[Y(0,1)\mid L_0 = l_0] / E[Y(0,0)\mid L_0 = l_0].$

	\begin{figure}
		\centering
		\begin{tikzpicture}[scale=2]
		\begin{scope}
		\draw[->] (-0.05,0) -- (1.5,0);
		\draw[->] (0,-0.05) -- (0,1.5);
		\node[anchor=base] at (1,-0.2){$1$}; 
		\node at (-0.15,1){\rotatebox{90}{$1$}};  
		\node[anchor=base] at (0,-0.2){$0$}; 
		\node at (-0.15,0){\rotatebox{90}{$0$}};  
		\node[scale=0.5, xshift=2mm] at (1.5,-0.1){\rotatebox{0}{$E[Y(0,0)\,|\, A_0\!=\!0,L_1\!=\!l_1]$}};  
		\node[scale=0.5,yshift=2mm] at (-0.1,1.5){\rotatebox{90}{$E[Y(0,1)\,|\,A_0\!=\!0,L_1\!=\!l_1]$}};  
		\node at (1.3,1.3){(a)};
		\draw[thick] (0,0) rectangle (1,1);
		\draw[thick,red] (0.8,0) -- (1,1-0.8);
		\draw[thick,red,dashed] (0.6,0) -- (1,1-0.6);
		\node (a1) at (0.6,-0.05){};
		\node (b1) at (1,-0.2){};
		\end{scope}
		\begin{scope}[yshift=-2.5cm]
		\draw[->] (-0.05,0) -- (1.5,0);
		\draw[->] (0,-0.05) -- (0,1.5);
		\node[anchor=base] at (1,-0.2){$1$}; 
		\node at (-0.15,1){\rotatebox{90}{$1$}};  
		\node[anchor=base] at (0,-0.2){$0$}; 
		\node at (-0.15,0){\rotatebox{90}{$0$}};  
		\node[scale=0.5] at (1.5,-0.1){\rotatebox{0}{$E[Y(0,0)]$}};  
		\node[scale=0.5] at (-0.1,1.5){\rotatebox{90}{$E[Y(1,0)]$}};  
		\node at (1.3,1.3){(b)};
		\draw[thick] (0,0) rectangle (1,1);
		\fill[fill=red!20!white,opacity=0.8] (0.6,0) rectangle (1,1);
		\fill[fill=blue!20!white,opacity=0.8] (0,0.45) rectangle (1,1);
		\draw[pattern=dots, pattern color= purple, ,draw opacity=0] (0.6,0.45) rectangle (1,1);
		\node (a2) at (0.6,1){};
		\node (b2) at (1,1){};
		\node (c2) at (1,0.45){};
		\node (d2) at (1,1){};
		\draw[thick,black!20!white] (0,0.6) -- (1-0.6,1);
		\draw[thick,black!20!white] (0.8,0) -- (1,1-0.8);
		\end{scope}
		\begin{scope}[xshift=2.5cm,yshift=-2.5cm]
		\draw[->] (-0.05,0) -- (1.5,0);
		\draw[->] (0,-0.05) -- (0,1.5);
		\node[anchor=base] at (1,-0.2){$1$}; 
		\node at (-0.15,1){\rotatebox{90}{$1$}};  
		\node[anchor=base] at (0,-0.2){$0$}; 
		\node at (-0.15,0){\rotatebox{90}{$0$}};  
		\node[scale=0.5, xshift=2mm] at (1.5,-0.1){\rotatebox{0}{$E[Y(1,1)\,|\, A_0\!=\!1,L_1\!=\!l_1]$}};  
		\node[scale=0.5,yshift=2mm] at (-0.1,1.5){\rotatebox{90}{$E[Y(1,0)\,|\,A_0\!=\!1,L_1\!=\!l_1]$}};  
		\draw[thick] (0,0) rectangle (1,1);
		\draw[thick,blue] (0,0.7) -- (1-0.7,1);
		\draw[thick,dashed,blue] (0,0.45) -- (1-0.45,1);
		\node (c1) at (-0.05,0.45){};
		\node (d1) at (-0.2,1){};
		\node at (1.3,1.3){(c)};
		\end{scope}
		\draw[thick,dotted,red,->] (a1) -- (a2);
		\draw[thick,dotted,red,->] (b1) -- (b2);
		\draw[thick,dotted,blue,->] (c1) -- (c2);
		\draw[thick,dotted,blue,->] (d1) -- (d2);
		\end{tikzpicture}
		\caption{Illustration of variation dependence of additive blip functions. For simplicity, we suppress the dependence on the baseline covariates $L_0$. The lines at 45 degrees in (a) give values for the second
			stage additive blip quantities $E[Y(0,1) - Y(0,0) \,|\, A_0=0,L_1=l_1], l_1=0,1$. 
			Similarly (c) shows the second stage blips with $A_0=1$. (b) shows the first stage blip $E[Y(1,0) - Y(0,0)]$. The dotted arrows from (a), (c) to (b) show restrictions placed on the the first stage quantities $E[Y(a_0,0)], a_0=0,1$, due to restrictions on the second stage quantities $E[Y(a_0,0)\mid A_0=a_0,L_1=l_1], a_0=0,1$. 
			The dotted area  in (b) shows the feasible region of $(E[Y(0,0)], E[Y(1,0)])$ given the second stage quantities depicted in (a) and (c), and the two gray lines at 45 degrees are two contour lines for the first stage blip $E[Y(1,0)-Y(0,0)]$  that are not feasible, i.e. that do no intersect with the dotted region. See the text for more explanation.}
		\label{fig:thomas-rd}
	\end{figure}

	\begin{figure}
		\centering
		\begin{tikzpicture}[scale=2]
		\begin{scope}
		\draw[->] (-0.05,0) -- (1.5,0);
		\draw[->] (0,-0.05) -- (0,1.5);
		\node[anchor=base] at (1,-0.2){$1$}; 
		\node at (-0.15,1){\rotatebox{90}{$1$}};  
		\node[anchor=base] at (0,-0.2){$0$}; 
		\node at (-0.15,0){\rotatebox{90}{$0$}};  
		\node[scale=0.5, xshift=2mm] at (1.5,-0.1){\rotatebox{0}{$E[Y(0,0)\,|\, A_0\!=\!0,L_1\!=\!l_1]$}};  
		\node[scale=0.5,yshift=2mm] at (-0.1,1.5){\rotatebox{90}{$E[Y(0,1)\,|\,A_0\!=\!0,L_1\!=\!l_1]$}};  
		\node at (1.3,1.3){(a)};
		\draw[thick] (0,0) rectangle (1,1);
		\draw[thick,red] (0,0) -- (1-0.8,1);
		\draw[thick,red,dashed] (0,0) -- (1-0.6,1);
		\node (a1) at (0,-0.2){};
		\node (b1) at (1-0.6,1){};
		\end{scope}
		\begin{scope}[yshift=-2.5cm]
		\draw[->] (-0.05,0) -- (1.5,0);
		\draw[->] (0,-0.05) -- (0,1.5);
		\node[anchor=base] at (1,-0.2){$1$}; 
		\node at (-0.15,1){\rotatebox{90}{$1$}};  
		\node[anchor=base] at (0,-0.2){$0$}; 
		\node at (-0.15,0){\rotatebox{90}{$0$}};  
		\node[scale=0.5] at (1.5,-0.1){\rotatebox{0}{$E[Y(0,0)]$}};  
		\node[scale=0.5] at (-0.1,1.5){\rotatebox{90}{$E[Y(1,0)]$}};  
		\node at (1.3,1.3){(b)};
		\draw[thick] (0,0) rectangle (1,1);
		\fill[fill=red!20!white,opacity=0.8] (0,0) rectangle (1-0.6,1);
		\fill[fill=blue!20!white,opacity=0.8] (0,0) rectangle (1,0.45);
		\draw[pattern=dots, pattern color= purple, ,draw opacity=0] (0,0) rectangle (0.4,0.45);
		\node (a2) at (0,1.5){};
		\node (b2) at (1-0.6,1){};
		\node (c2) at (1.5,0){};
		\node (d2) at (1,0.45){};
		\end{scope}
		\begin{scope}[xshift=2.5cm,yshift=-2.5cm]
		\draw[->] (-0.05,0) -- (1.5,0);
		\draw[->] (0,-0.05) -- (0,1.5);
		\node[anchor=base] at (1,-0.2){$1$}; 
		\node at (-0.15,1){\rotatebox{90}{$1$}};  
		\node[anchor=base] at (0,-0.2){$0$}; 
		\node at (-0.15,0){\rotatebox{90}{$0$}};  
		\node[scale=0.5, xshift=2mm] at (1.5,-0.1){\rotatebox{0}{$E[Y(1,1)\,|\, A_0\!=\!1,L_1\!=\!l_1]$}};  
		\node[scale=0.5,yshift=2mm] at (-0.1,1.5){\rotatebox{90}{$E[Y(1,0)\,|\,A_0\!=\!1,L_1\!=\!l_1]$}};  
		\draw[thick] (0,0) rectangle (1,1);
		\draw[thick,blue] (0,0) -- (1,0.25);
		\draw[thick,dashed,blue] (0,0) -- (1,0.45);
		\node (c1) at (-0.2,0){};
		\node (d1) at (1,0.45){};
		\node at (1.3,1.3){(c)};
		\end{scope}
		\draw[thick,dotted,red,->] (a1) -- (a2);
		\draw[thick,dotted,red,->] (b1) -- (b2);
		\draw[thick,dotted,blue,->] (c1) -- (c2);
		\draw[thick,dotted,blue,->] (d1) -- (d2);
		\end{tikzpicture}
		\caption{Illustration of variation independence of multiplicative blip functions. For simplicity, we suppress the dependence on the baseline covariates $L_0$. The lines through the origin in (a) give values for the second
			stage additive blip quantities  $E[Y(0,1)  \,|\, A_0=0,L_1=l_1]/E[Y(0,0)  \,|\, A_0=0,L_1=l_1], l_1=0,1$. 
			Similarly (c) shows the second stage blips with $A_0=1$.  In (b) the intersection of the shaded regions indicates possible values for
			$(E[Y(0,0)],  E[Y(1,0)])$. Since this set contains the origin $(0,0),$ it will intersect any contour line for the multiplicative first stage blip.
			Thus the first and second stage multiplicative blips are variation independent.}
		\label{fig:thomas-rr}
	\end{figure}

\section{Extension to general time-varying treatments and confounders}
\label{sec:extension-general-treatment}

In this section, we discuss how the proposed parameterization in Section \ref{subsec:2time} may be extended to accommodate general time-varying treatments and confounders. These extensions are based on similar results in a recent paper by \cite{yin2022multiplicative}, so we refer interested readers there for a detailed discussion and proofs.  To focus on the main ideas, we only discuss the extension in the case with $K=1.$ The general case discussed in Section \ref{subsec:general} may be extended in a similar way and hence omitted.

\subsection{Extension to categorical time-varying treatments and confounders}

Suppose that the treatment variable $A_0, A_1$ have $K_A+1$ levels of treatments, denoted by $0,1,\ldots, K_A$, and the time-varying covariate $L_1$ has $K_L +1$ levels of treatments, denoted by $0,1,\ldots, K_L.$ Note that for notational simplicity, we have assumed that $A_0$ and $A_1$ take the same number of levels.

	Suppose that the sequential ignorability assumption \eqref{eqn:seq_igno_2time} holds. For notational simplicity, suppose that the value for baseline covariates $L_0$ is fixed, and we suppress the dependence on $L_0$ in the expressions below.	 Let $\mathcal{M}$ denote the models consisting of the SNMMs on
\begin{align*}
    \theta_0(a_0) = \dfrac{E[Y(a_0,0)]}{E[Y(0,0)]}, \quad a_0 = 1,2, \ldots, K_A;
\end{align*}
\begin{align*}
    \theta_1(a_0, l_1, a_1) = \dfrac{E[Y(a_0,a_1)| A_0 = a_0, L_1 = l_1]}{E[Y(a_0,0)| A_0 = a_0, L_1 = l_1]} = \dfrac{E[Y| A_0 = a_0, L_1 = l_1, A_1 = a_1]}{E[Y| A_0 = a_0, L_1 = l_1, A_1 = 0]}, \\
    \quad a_0 = 0,1, \ldots, K_A, l_1 = 0,1, \ldots, K_L, a_1 = 1,2, \ldots, K_A
\end{align*}
and models on the following nuisance parameters:
\begin{align*}
\phi(a_0, l_1)  = \dfrac{E[Y(a_0,0)| A_0 = a_0, L_1 = l_1]}{E[Y(a_0,0)| A_0 = a_0, L_1 = 0]} = \dfrac{E[Y| A_0 = a_0, L_1 = l_1, A_1 = 0]}{E[Y| A_0 = a_0, L_1 = 0, A_1 = 0]}, \\
    \quad a_0 = 0,1, \ldots, K_A, l_1 = 1,2, \ldots, K_L;
\end{align*}
\begin{align*}
    \text{gop} &= \dfrac{\prod\limits_{a_0=0,\ldots, K_A}\prod\limits_{l_1=0,\ldots,K_L}\prod\limits_{a_1=0,\ldots, K_A} E[Y\mid A_0=a_0, L_1=l_1, A_1=a_1] }{\prod\limits_{a_0=0,\ldots, K_A}\prod\limits_{l_1=0,\ldots,K_L}\prod\limits_{a_1=0,\ldots, K_A} (1-E[Y\mid A_0=a_0, L_1=l_1, A_1=a_1])}, 
\end{align*}
\begin{align*}
\eta^*(a_0,l_1) = \dfrac{\sum\limits_{l^* = 0}^{l_1} P(L_1 = l_1|A_0 = a_0)}{\sum\limits_{l^* = 0}^{l_1+1} P(L_1 = l_1|A_0 = a_0)}, \quad a_0 = 0,1, \ldots, K_A, l_1 = 0,1, \ldots, K_L-1.
\end{align*}
Then the map given by 
		\begin{flalign*}
		&	(\theta_0(a_0), \theta_1(a_0,l_1,a_1), \phi(a_0,l_1), \text{gop},\eta^*(a_0,l_1)) \rightarrow \\
		&\quad \quad \quad \quad	 (P(Y=1\mid A_0=a_0, L_1=l_1, A_1=a_1), \eta(a_0,l_1))
		\numberthis	\label{map:one-to-one:sup}
		\end{flalign*}				
		is a bijection from $(\mathbb{R}^{+})^{(K_A+1)^2 (K_L+1)} \times (0,1)^{(K_A+1)K_L}$ to $(0,1)^{ (K_A+1)^2 (K_L+1) + (K_A+1)K_L  }$; note in \eqref{map:one-to-one:sup} we have omitted the range of values for $a_0, l_1, a_1$ for notational simplicity. 	Furthermore, the models specifying $\mathcal{M}$ are variation independent of each other. 

\subsection{Extension to continuous time-varying treatments and confounders}

Suppose now that the treatment variables $A_0, A_1$ are continuous, taking values in $[a_{min}, a_{max}]$. Without loss of generality, assume that $a_{min} \leq 0 \leq a_{max}.$ Suppose also that  the time-varying covariate $L_1$ is continuous, taking values in $[l_{min}, l_{max}]$. We also assume that $A_0, L_1, A_1$ are coded in such a way that the following monotonicity assumption holds.
\begin{assumption}
\label{assumption:monotone}
$E[Y\mid A_0, L_1, A_1]$ is monotone in $A_0, L_1, A_1.$ 
\end{assumption}

Suppose that the sequential ignorability assumption \eqref{eqn:seq_igno_2time} holds. For notational simplicity, suppose that the value for baseline covariates $L_0$ is fixed, and we suppress the dependence on $L_0$ in the expressions below.	 Let $\mathcal{M}$ denote the models specified by the SNMMs:
\begin{align*}
   \theta_0(a_0) = \dfrac{E[Y(a_0,0)]}{E[Y(0,0)]}
\end{align*}
\begin{align*}
    \theta_1(a_0, l_1, a_1) = \dfrac{E[Y(a_0,a_1)| A_0 = a_0, L_1 = l_1]}{E[Y(a_0,0)| A_0 = a_0, L_1 = l_1]} = \dfrac{E[Y| A_0 = a_0, L_1 = l_1, A_1 = a_1]}{E[Y| A_0 = a_0, L_1 = l_1, A_1 = 0]}, \\
\end{align*}
and models on the following nuisance parameters:
\begin{align*}
\phi(a_0, l_1)  = \dfrac{E[Y(a_0,0)| A_0 = a_0, L_1 = l_1]}{E[Y(a_0,0)| A_0 = a_0, L_1 = 0]} = \dfrac{E[Y| A_0 = a_0, L_1 = l_1, A_1 = 0]}{E[Y| A_0 = a_0, L_1 = 0, A_1 = 0]}; 
\end{align*}
\begin{align*}
   	\text{gop} &\equiv \dfrac{\prod\limits_{a_0=a_{min},a_{max}}\prod\limits_{l_1=l_{min},l_{max}}\prod\limits_{a_1=a_{min},a_{max}} E[Y\mid A_0=a_0, L_1=l_1, A_1=a_1] }{\prod\limits_{a_0=a_{min},a_{max}}\prod\limits_{l_1=l_{min},l_{max}}\prod\limits_{a_1=a_{min},a_{max}} (1-E[Y\mid A_0=a_0, L_1=l_1, A_1=a_1])}; 
\end{align*}
\begin{align*}
	\eta(a_0) &\equiv f(L_1 \mid A_0=a_0);
\end{align*}
here $f(L_1 \mid A_0=a_0)$ denotes the conditional density of $L_1$ given $A_0.$ 
Then given $\eta(a_0),$ the map given by 
		\begin{flalign*}
			(\theta_0(a_0), \theta_1(a_0,l_1,a_1), \phi(a_0,l_1), \text{gop}) \rightarrow 
		 (P(Y=1\mid A_0=a_0, L_1=l_1, A_1=a_1))
		\numberthis	\label{map:one-to-one:sup2}
		\end{flalign*}				
		is a bijection between their domains.
Furthermore, models in $\mathcal{M}$ are variation independent of each other.


It remains a challenge to specify models in $\mathcal{M}$ so that Assumption \ref{assumption:monotone} holds. Note that simple models on $\theta_1(a_0,l_1,a_1)$, such as 
$$
    \log \theta_1(a_0,l_1,a_1) = \alpha_0  + \alpha_1 a_0 + \alpha_2 l_1 + \alpha_3 a_1
$$
imply that $E[Y\mid A_0=a_0, L_1=l_1, A_1=a_1]$ is monotone in $a_1.$ Although one can similarly specify 
$$
    \log \phi(a_0,l_1) = \beta_0 + \beta_1 a_0 + \beta_2 l_1,
$$
it only implies that $E[Y\mid A_0=a_0, L_1=l_1, A_1=0]$ is monotone in $l_1$. In other words, there is no guarantee that $E[Y\mid A_0=a_0, L_1=l_1, A_1=a_1]$ is monotone in $l_1$ for $a_1\neq 0.$ Similar problems arise for specifying models so that $E[Y\mid A_0=a_0, L_1=l_1, A_1=a_1]$ is monotone in $a_0.$ We leave specification of models in $\mathcal{M}$ as a future research topic.

\section{Proof of Theorem \ref{thm:main}}
	\label{appendix:proof}
	\begin{proof} We suppress dependence on $L_0$ throughout this proof.
		Let $p_{a_0 l_1 a_1} = E[Y\mid A_0=a_0, L_1=l_1, A_1=a_1]$, 
		We first note that due to the g-formula \citep{robins1986new}, 
		\begin{flalign*}
		\theta_0 &= \dfrac{\eta(1) p_{110} + (1-\eta(1)) p_{100}}{\eta(0) p_{010} + (1-\eta(0)) p_{000}} \\
		&=  \dfrac{\eta(1) \dfrac{p_{110}}{p_{100}} + 1-\eta(1)} {\eta(0) \dfrac{p_{010}}{p_{000}} + 1-\eta(0)} \times \dfrac{p_{100}}{p_{000}} \\
		&= \dfrac{\eta(1) \phi(1) + 1-\eta(1)}{\eta(0) \phi(0) + 1-\eta(0)} \times \dfrac{p_{100}}{p_{000}}, \numberthis\label{eqn:theta0}
		\end{flalign*}
		hence 
		\begin{small}
			\begin{equation}
			(\theta_0,\theta_1(1,1),\ldots,\theta_1(0,0),\phi(0),\phi(1),\text{gop},\eta(0),\eta(1)) \rightarrow \left( \dfrac{p_{100}}{p_{000}}, \dfrac{p_{111}}{p_{110}}, \dfrac{p_{101}}{p_{100}},\dfrac{p_{011}}{p_{010}},\dfrac{p_{001}}{p_{000}},\dfrac{p_{010}}{p_{000}},\dfrac{p_{110}}{p_{100}}, \text{gop}, \eta(0),\eta(1)   \right)\label{eq:automorphism-one}
			\end{equation}
		\end{small}
		is an automorphism on $(\mathbb{R}^{+})^8 \times (0,1)^2$; note that in simple terms, the mapping (\ref{eq:automorphism-one}) 
		simply replaces $\theta_0$ with 
		$\dfrac{p_{100}}{p_{000}}$.
		Applying simple algebra to the RHS of (\ref{eq:automorphism-one}), we have that 
		\begin{small}
			\begin{equation}
			(\theta_0,\theta_1(1,1),\ldots,\theta_1(0,0),\phi(0),\phi(1),\text{gop},\eta(0),\eta(1)) \rightarrow \left( \dfrac{p_{111}}{p_{000}}, \dfrac{p_{110}}{p_{000}}, \dfrac{p_{101}}{p_{000}},\dfrac{p_{100}}{p_{000}},\dfrac{p_{011}}{p_{000}},\dfrac{p_{010}}{p_{000}},\dfrac{p_{001}}{p_{000}}, \text{gop}, \eta(0),\eta(1)   \right)
			\label{eq:automorphism-two} \end{equation}
		\end{small}%
		is also  an automorphism on $(\mathbb{R}^{+})^8 \times (0,1)^2$. Note that the numerators of the first seven terms on the RHS of  (\ref{eq:automorphism-two})
		are simply consecutive  terms of $p_{a_0, l_1, a_1}$ with binary translation on the indexes. Now what is left to show is that given $\eta(0), \eta(1)$,
		\begin{equation}
		\label{eqn:bijection}
		(p_{a_0 l_1 a_1}, a_0,l_1,a_1\in\{0,1\})  \rightarrow \left( \dfrac{p_{111}}{p_{000}}, \dfrac{p_{110}}{p_{000}}, \dfrac{p_{101}}{p_{000}},\dfrac{p_{100}}{p_{000}},\dfrac{p_{011}}{p_{000}},\dfrac{p_{010}}{p_{000}},\dfrac{p_{001}}{p_{000}}, \text{gop} \right) 
		\end{equation}
		is a bijection from $(0,1)^8$ to $(\mathbb{R}^{+})^8$.
		
		To show that \eqref{eqn:bijection} is a bijection, let $\bm c = (c_1,\ldots,c_8)$ be a  vector in $(\mathbb{R}^{+})^8$.
		We need to show that there is one and only one $\bm p = (p_{a_0 l_1 a_1}, a_0,l_1,a_1\in\{0,1\}) \in (0,1)^8$ such that
		\begin{equation}
		\label{eqn:mapping}
		\left( \dfrac{p_{111}}{p_{000}}, \dfrac{p_{110}}{p_{000}}, \dfrac{p_{101}}{p_{000}},\dfrac{p_{100}}{p_{000}},\dfrac{p_{011}}{p_{000}},\dfrac{p_{010}}{p_{000}},\dfrac{p_{001}}{p_{000}} \right)  = (c_7,\ldots,c_1)
		\end{equation}	
		and 
		\begin{equation}
		\label{eqn:c8}
		\text{gop} = c_8.
		\end{equation}
		Let $r_{a_0,l_1,a_1} =   c_{4{a_0} + 2{l_1} + a_1}$ if $4{a_0} + 2{l_1} + a_1\geq 1$, $r_{0,0,0} = 1,$
		$ k_{a_0l_1a_1}  = r_{a_0l_1a_1} / \max\limits_{a_0,l_1,a_1} r_{a_0,l_1,a_1}$ and  $\bm k = \{k_{a_0l_1a_1}:a_0,l_1,a_1\in\{0,1\}\}$. It is easy to see that   $k_{max} = \max\limits_{a_0,l_1,a_1} k_{a_0l_1a_1} = 1$ and $$ 	
		\left\{ \bm p \in (0,1)^8: \eqref{eqn:mapping}  \text{ holds} \right\}  = \left\{ p_{max} \bm k: 0 <p_{max} < 1 \right\}.
		$$  Thus for any $\bf p$ such that \eqref{eqn:mapping} holds, the constraint \eqref{eqn:c8} may equivalently be expressed as
		\begin{equation*}
		\log(\text{gop}) = \sum\limits_{i=(a_0, l_1, a_1)} \log\left( k_i \right) + 8 \log(p_{max}) - \sum\limits_{i=(a_0, l_1, a_1)} \log(1-k_ip_{max}) = \log(c_8).
		\end{equation*}
		Let $g(p_{max}) = \sum\limits_{i=a_0, l_1, a_1} \log\left( k_i \right) + 8 \log(p_{max}) - \sum\limits_{i=a_0, l_1, a_1} \log(1-k_ip_{max}) - \log(c_8)$. We claim that $g(p_{max})$ has one and only one solution in the interval (0,1) so that there is one and only one $\bm p = p_{max} \bm k \in (0,1)^8$ such that  both \eqref{eqn:mapping} and \eqref{eqn:c8} hold. 
		
		We now prove our claim. Note that  
		\begin{flalign*}
		\dfrac{\partial}{\partial p_{max}}g(p_{max}) &= 	 \dfrac{8}{p_{max}} + \sum\limits_{i=a_0, l_1, a_1} \dfrac{k_i}{1-k_ip_{max}} > 0
		\end{flalign*}
		since, by construction, $k_ip_{max} < 1$ for all $i$.
		Hence $g(p_{max}) $ is monotone on $(0,1)$. Furthermore, 
		$\lim\limits_{p_{max}\rightarrow 0} g(p_{max}) = -\infty$, $\lim\limits_{p_{max}\rightarrow 1} g(p_{max}) = +\infty$ and $g(p_{max}) $ is also continuous on $(0,1)$. As a result, $g(p_{max})$ has one and only one root in $(0,1)$. 
		
		Lastly, since the domain of $\mathcal{M}$, i.e. $(\mathbb{R}^+)^8 \times (0,1)^2$ is the Cartesian product of the marginal domains of models in $\mathcal{M}$, models in $\mathcal{M}$ are variation independent of each other.

		We have hence finished the proof.
	\end{proof}

\section{Algorithm for computing the mapping \eqref{eqn:mapping_bivariate}}
\label{sec:algorithm}

Algorithm \ref{alg:estimate_prob} details the steps to compute the mapping \eqref{eqn:mapping_bivariate}.

		\begin{algorithm}
 			\caption{\ \ An algorithm for computing $E[Y\mid \bar{A}_K=\bar{a}_K, \bar{L}_K=\bar{l}_K]$ from $(\bm\theta, \bm\phi, \text{GOP}(l_0),\bm \eta)$}
		\label{alg:estimate_prob}
		\begin{enumerate}
			\item  {\bf For} $k = 0,\ldots,K$, compute \\[5pt]
			$ \dfrac{E[Y\mid \bar{A}_K = (\bar{a}_{k-1},1, \bm 0), \bar{L}_K = (\bar{l}_k,\bm 0)]}{E[Y\mid \bar{A}_K = (\bar{a}_{k-1}, 0, \bm 0), \bar{L}_K = (\bar{l}_k,\bm 0)]}$ 
			and $ \dfrac{E[Y\mid \bar{A}_K = (\bar{a}_k, \bm 0), \bar{L}_K = (\bar{l}_k,1,\bm 0)]}{E[Y\mid \bar{A}_K = (\bar{a}_{k}, \bm 0), \bar{L}_K = (\bar{l}_k,0, \bm 0)]}$\\[5pt]
			using equations \eqref{eqn:1} and \eqref{eqn:2}. \\[-4pt]
			\item {\bf For} $k = 0,\ldots,K$, compute \\[5pt]
			 $\dfrac{E[Y\mid \bar{A}_K = (\bar{a}_{k-1},1, \bm 0), \bar{L}_K = (\bar{l}_k,\bm 0)]}{E[Y\mid \bar{A}_K = \bm 0, \bar{L}_K = (l_0,\bm 0)]}$ and $\dfrac{E[Y\mid \bar{A}_K = (\bar{a}_k, \bm 0), \bar{L}_K = (\bar{l}_k,1,\bm 0)]}{E[Y\mid \bar{A}_K = \bm 0, \bar{L}_K =  (l_0,\bm 0)]}$ sequentially by
			\begin{flalign*}
			\footnotesize \dfrac{E[Y\mid \bar{A}_K = (\bar{a}_{k-1},1, \bm 0), \bar{L}_K = (\bar{l}_k,\bm 0)]}{E[Y\mid \bar{A}_K = \bm 0, \bar{L}_K = (l_0,\bm 0)]} &=  \footnotesize   \dfrac{E[Y\mid \bar{A}_K = (\bar{a}_{k-1},1, \bm 0), \bar{L}_K = (\bar{l}_k,\bm 0)]}{E[Y\mid \bar{A}_K = (\bar{a}_{k-1}, 0, \bm 0), \bar{L}_K = (\bar{l}_k,\bm 0)]} \times \dfrac{E[Y\mid \bar{A}_K = (\bar{a}_{k-1},\bm 0, \bm 0), \bar{L}_K = (\bar{l}_k,\bm 0)]}{E[Y\mid \bar{A}_K = \bm 0, \bar{L}_K = (l_0,\bm 0)]}; \\
			\footnotesize \dfrac{E[Y\mid \bar{A}_K = (\bar{a}_k, \bm 0), \bar{L}_K = (\bar{l}_k,1,\bm 0)]}{E[Y\mid \bar{A}_K = \bm 0, \bar{L}_K = (l_0,\bm 0)]} &= \footnotesize \dfrac{E[Y\mid \bar{A}_K = (\bar{a}_k, \bm 0), \bar{L}_K = (\bar{l}_k,1,\bm 0)]}{E[Y\mid \bar{A}_K = (\bar{a}_{k}, \bm 0), \bar{L}_K = (\bar{l}_k,0, \bm 0)]}  \times \dfrac{E[Y\mid \bar{A}_K = (\bar{a}_{k}, \bm 0), \bar{L}_K = (\bar{l}_k,\bm 0)]}{E[Y\mid \bar{A}_K =\bm 0, \bar{L}_K = (l_0,\bm 0)]}.
			\end{flalign*}\\[-4pt]
			\item Compute
			$$
			r_{\text{max}}(l_0) = \max\limits_{\bar{a}_K, l_1,\ldots ,l_K}  r_{\bar{a}_K, l_1,\ldots ,l_K}(l_0),
			$$
			where $r_{\bar{a}_K, l_1,\ldots ,l_K}(l_0) = E[Y\,|\, \bar{A}_K = \bar{a}_K, \bar{L}_K = \bar{l}_K] / E[Y\,|\, \bar{A}_K = \bm 0, \bar{L}_K = (l_0,\bm 0)]$ is computed in Step 2.\\[-4pt]
			\item Compute
			$$
			k_{\bar{a}_K, l_1,\ldots ,l_K}(l_0) \equiv 		\dfrac{	r_{\bar{a}_K, l_1,\ldots ,l_K}(l_0)}{ r_{\text{max}}(l_0)}.
			$$
			\\[-4pt]
			\item Let $p_{\bar{a}_K, \bar{l}_K} = E[Y\mid \bar{A}_K = \bar{a}_K, \bar{L}_K = \bar{l}_K]$, and 
			suppressing dependence on $l_0$, for $x\in (0,1),$ let
			$$
			g(x) = \sum\limits_{i\in\mathcal{I}} \log(k_i) + 2^{2K+1} \log(x)  - \sum\limits_{i\in \mathcal{I}}  \log(1-k_i x) - \log(GOP),		 $$
			where $\mathcal{I}=\{(\bar{a}_K, (l_1,\ldots ,l_K)): \bar{a}_K \in \{0,1\}^{K+1}, (l_1,\ldots ,l_K)\in\{0,1\}^K\}.$
			Find  the unique root of $g(x)$ in the interval $(0,1)$.  Set $p_{\text{max}}(l_0)$ to be this root.\\[-4pt]
			%
			%
			\item  Compute 
			$$
			E[Y\mid \bar{A}_K = \bar{a}_K, \bar{L}_K = \bar{l}_K] = k_{\bar{a}_K, l_1,\ldots,{l}_K}(l_0) \times p_{\text{max}}(l_0).
			$$
		\end{enumerate}
	\end{algorithm}

\section{Proof of Theorem \ref{thm:main_general}}
	\label{section:proof2}
	\begin{proof}
		We follow the outline provided in Algorithm \ref{alg:estimate_prob}. First, note that
		\vspace{-0.5cm}
		\begin{flalign*}
\MoveEqLeft[0]{\hspace{-2.50cm}\quad \ \ \theta_k(\bar{a}_{k-1}, \bar{l}_k)}\\
		&\hspace{-2.00cm}= \dfrac{E[Y(\bar{a}_{k-1},1, \bm 0) \mid \bar{A}_k=(\bar{a}_{k-1},1), \bar{L}_k=\bar{l}_k]}{E[Y(\bar{a}_{k-1}, 0, \bm 0) \mid \bar{A}_k=(\bar{a}_{k-1},1), \bar{L}_k=\bar{l}_k]} \\ 
		&\hspace{-2.00cm}= \dfrac{E[Y(\bar{a}_{k-1},1, \bm 0) \mid \bar{A}_k=(\bar{a}_{k-1},1), \bar{L}_k=\bar{l}_k]}{E[Y(\bar{a}_{k-1}, 0, \bm 0) \mid \bar{A}_k=(\bar{a}_{k-1}, 0), \bar{L}_k=\bar{l}_k]} \text{\quad (by sequential ignorability)} \\ 
		&\hspace{-2.00cm}= \small	\dfrac{		\eta((\bar{a}_{k-1},1), \bar{l}_k) E[Y(\bar{a}_{k-1},1, \bm 0) \mid \bar{A}_k\!=\!(\bar{a}_{k-1},1), \bar{L}_{k+1} = (\bar{l}_k,1)]  + (1-	\eta((\bar{a}_{k-1},1), \bar{l}_k)) E[Y(\bar{a}_{k-1},1, \bm 0) \mid \bar{A}_k\!=\!(\bar{a}_{k-1},1), \bar{L}_{k+1}\!=\!(\bar{l}_k,0)] }   
		{ \eta((\bar{a}_{k-1}, 0), \bar{l}_k) E[Y(\bar{a}_{k-1}, 0, \bm 0) \mid \bar{A}_k\!=\!(\bar{a}_{k-1}, 0), \bar{L}_{k+1}\!=\! (\bar{l}_k,1)]  + (1-	\eta((\bar{a}_{k-1}, 0), \bar{l}_k)) E[Y(\bar{a}_{k-1}, 0, \bm 0) \mid \bar{A}_k\!=\!(\bar{a}_{k-1}, 0), \bar{L}_{k+1}\! =\! (\bar{l}_k,0)]} \\
		&\text{\hspace{-2.00cm} (by law of total expectation)} \\
		&\hspace{-2.00cm}= 	\dfrac{	\eta((\bar{a}_{k-1},1), \bar{l}_k)\phi((\bar{a}_{k-1},1), \bar{l}_k)  + 1-\eta((\bar{a}_{k-1},1), \bar{l}_k)  }   
		{\eta((\bar{a}_{k-1}, 0), \bar{l}_k) \phi((\bar{a}_{k-1}, 0), \bar{l}_k)   + 1-\eta((\bar{a}_{k-1}, 0), \bar{l}_k) } \times \dfrac{E[Y(\bar{a}_{k-1},1, \bm 0) \mid \bar{A}_k=(\bar{a}_{k-1},1), \bar{L}_{k+1} = (\bar{l}_k,0)]}{E[Y(\bar{a}_{k-1}, 0, \bm 0) \mid \bar{A}_k=(\bar{a}_{k-1}, 0), \bar{L}_{k+1} = (\bar{l}_k,0)]} \\
		&\hspace{-2.00cm}= \dfrac{	\eta((\bar{a}_{k-1},1), \bar{l}_k) \phi((\bar{a}_{k-1},1), \bar{l}_k)  + 1-\eta((\bar{a}_{k-1},1), \bar{l}_k)  }   
		{\eta((\bar{a}_{k-1}, 0), \bar{l}_k) \phi((\bar{a}_{k-1}, 0), \bar{l}_k)   + 1-\eta((\bar{a}_{k-1}, 0), \bar{l}_k) } \times \dfrac{E[Y(\bar{a}_{k-1},1, \bm 0) \mid \bar{A}_{k+1}=(\bar{a}_{k-1},1,0), \bar{L}_{k+1} = (\bar{l}_k,0)]}{E[Y(\bar{a}_{k-1}, 0, \bm 0) \mid \bar{A}_{k+1}=(\bar{a}_{k-1}, 0,0), \bar{L}_{k+1} = (\bar{l}_k,0)]} \\ 
		&\text{ \hspace{-2.00cm} (by sequential ignorability)} \\
		&\hspace{-2.00cm}= \cdots \\
		&\hspace{-2.00cm}= \footnotesize \prod\limits_{m=k}^{K-1}  \dfrac{	\eta((\bar{a}_{k-1},1,\vec{0}_{m-k}), (\bar{l}_k, \vec{0}_{m-k})) \phi((\bar{a}_{k-1},1,\vec{0}_{m-k}), (\bar{l}_k, \vec{0}_{m-k}))  + 1-\eta((\bar{a}_{k-1},1,\vec{0}_{m-k}), (\bar{l}_k, \vec{0}_{m-k}))  }   
		{\eta((\bar{a}_{k-1}, 0, \vec{0}_{m-k}), (\bar{l}_{k}, \vec{0}_{m-k})) \phi((\bar{a}_{k-1}, 0, \vec{0}_{m-k}), (\bar{l}_{k}, \vec{0}_{m-k}))   + 1-\eta((\bar{a}_{k-1}, 0, \vec{0}_{m-k}), (\bar{l}_{k}, \vec{0}_{m-k})) }    \\
		&\hspace{-2.00cm}\quad \times\dfrac{E[Y\mid \bar{A}_K = (\bar{a}_{k-1},1, \bm 0), \bar{L}_K = (\bar{l}_k,\bm 0)]}{E[Y\mid \bar{A}_K = (\bar{a}_{k-1}, 0, \bm 0), \bar{L}_K = (\bar{l}_k,\bm 0)]}. \numberthis\label{eqn:1}
		\end{flalign*}
		Similarly, \vspace{-0.5cm}
		\begin{flalign*}
\MoveEqLeft[-5]{\hspace{-4.00cm}\phi(\bar{a}_k, \bar{l}_k)} \\
		&\hspace{-2.00cm}= \dfrac{E[Y(\bar{a}_k, \bm 0) \mid \bar{A}_{k+1} = (\bar{a}_k, 0), \bar{L}_{k+1} = (\bar{l}_k, 1)]}{E[Y(\bar{a}_k, \bm 0) \mid \bar{A}_{k+1} = (\bar{a}_k, 0), \bar{L}_{k+1} = (\bar{l}_k, 0)]} \text{\quad (by sequential ignorability)}\\  
		&\hspace{-2.00cm}=\footnotesize \dfrac{ \eta((\bar{a}_k, 0), (\bar{l}_k,1)) E[Y(\bar{a}_k, \bm 0) \mid \bar{A}_{k+1}\!=\!(\bar{a}_k, 0), \bar{L}_{k+2}\!=\! (\bar{l}_k, 1,1)] + (1-\eta((\bar{a}_k, 0), (\bar{l}_k,1))) E[Y(\bar{a}_k, \bm 0) \mid \bar{A}_{k+1}\!=\! (\bar{a}_k, 0), \bar{L}_{k+2}\!=\! (\bar{l}_k, 1,0)]}{ \eta((\bar{a}_k, 0), (\bar{l}_k,0)) E[Y(\bar{a}_k, \bm 0) \mid \bar{A}_{k+1}\!=\! (\bar{a}_k, 0), \bar{L}_{k+2} \!=\! (\bar{l}_k, 0,1)] + (1-\eta((\bar{a}_k, 0), (\bar{l}_k,0))) E[Y(\bar{a}_k, \bm 0) \mid \bar{A}_{k+1} \!=\! (\bar{a}_k, 0), \bar{L}_{k+2} \!=\! (\bar{l}_k, 0,0)]} \\
		&\hspace{-2.00cm}\text{ (by law of total expectation)} \\
		&\hspace{-2.00cm}= \dfrac{ \eta((\bar{a}_k, 0), (\bar{l}_k,1)) \phi((\bar{a}_k, 0),(\bar{l}_k,1)) + 1-\eta((\bar{a}_k, 0), (\bar{l}_k,1)) }{ \eta((\bar{a}_k, 0), (\bar{l}_k,0)) \phi((\bar{a}_k, 0),(\bar{l}_k,0))  + 1-\eta((\bar{a}_k, 0), (\bar{l}_k,0)) }  \times \dfrac{E[Y(\bar{a}_k,  \bm 0) \mid \bar{A}_{k+1} = (\bar{a}_k, 0), \bar{L}_{k+2} = (\bar{l}_k, 1,0)]}{E[Y(\bar{a}_k, \bm 0) \mid \bar{A}_{k+1} = (\bar{a}_k, 0), \bar{L}_{k+2} = (\bar{l}_k, 0,0)]} \\
		&\hspace{-2.00cm}=  \dfrac{ \eta((\bar{a}_k, 0), (\bar{l}_k,1)) \phi((\bar{a}_k, 0),(\bar{l}_k,1)) + 1-\eta((\bar{a}_k, 0), (\bar{l}_k,1)) }{ \eta((\bar{a}_k, 0), (\bar{l}_k,0)) \phi((\bar{a}_k, 0),(\bar{l}_k,0))  + 1-\eta((\bar{a}_k, 0), (\bar{l}_k,0)) } \times \dfrac{E[Y(\bar{a}_k, \bm 0) \mid \bar{A}_{k+2} = (\bar{a}_k, 0,0), \bar{L}_{k+2} = (\bar{l}_k, 1,0)]}{E[Y(\bar{a}_k, \bm 0) \mid \bar{A}_{k+2} = (\bar{a}_k, 0,0), \bar{L}_{k+2} = (\bar{l}_k, 0,0)]} \\
		&\hspace{-2.00cm}\text{  (by sequential ignorability)} \\
		&\hspace{-2.00cm}= \cdots \\
		&\hspace{-2.00cm}=  \prod\limits_{m=k}^{K-2}  \dfrac{\eta((\bar{a}_{k},0,\vec{0}_{m-k}), (\bar{l}_{k}, 1, \vec{0}_{m-k})) \phi((\bar{a}_{k},0,\vec{0}_{m-k}), (\bar{l}_{k}, 1, \vec{0}_{m-k}))  + 1-\eta((\bar{a}_{k},0,\vec{0}_{m-k}), (\bar{l}_{k}, 1, \vec{0}_{m-k}))  }   
		{\eta((\bar{a}_{k},0,\vec{0}_{m-k}), (\bar{l}_{k}, 0, \vec{0}_{m-k})) \phi((\bar{a}_{k},0,\vec{0}_{m-k}), (\bar{l}_{k},  0,\vec{0}_{m-k}))   + 1-\eta((\bar{a}_{k},0,\vec{0}_{m-k}), (\bar{l}_{k}, 0, \vec{0}_{m-k}))}   \\
		&\hspace{-2.00cm}\times \dfrac{E[Y\mid \bar{A}_K = (\bar{a}_k, \bm 0), \bar{L}_K = (\bar{l}_k,1,\bm 0)]}{E[Y\mid \bar{A}_K = (\bar{a}_{k}, \bm 0), \bar{L}_K = (\bar{l}_k, 0, \bm 0)]}.  \numberthis \label{eqn:2}
		\end{flalign*}

		Hence by recursive arguments, it is easy to see that for each value of $l_0$, the mapping
		\begin{equation}
		\label{eqn:auto}
		\small (\bm \theta, \bm \phi, GOP, \bm \eta) \rightarrow  \left(\dfrac{E[Y\mid \bar{A}_K = (\bar{a}_{k-1}, 1, \bm 0), \bar{L}_K = (\bar{l}_k,\bm 0)]}{E[Y\mid \bar{A}_K = (\bar{a}_{k-1}, 0, \bm 0), \bar{L}_K = (\bar{l}_k,\bm 0)]}, \dfrac{E[Y\mid \bar{A}_K = (\bar{a}_k, \bm 0), \bar{L}_K = (\bar{l}_k,1,\bm 0)]}{E[Y\mid \bar{A}_K = (\bar{a}_{k}, \bm 0), \bar{L}_K = (\bar{l}_k,0, \bm 0)]}, k=1,\ldots, K; GOP, \bm \eta \right)
		\end{equation}
		is  an automorphism on 
		$(\mathbb{R}^+)^{d_1} \times (0,1)^{d_2}$.
		Applying simple algebra to the RHS of (\ref{eqn:auto}) as in Step 2 of Algorithm \ref{alg:estimate_prob}, we have that for each value of $l_0$, 
		\begin{equation*}
		(\bm \theta, \bm \phi, GOP, \bm \eta) \rightarrow  \left(\dfrac{E[Y\mid \bar{A}_K = \bar{a}_{K}, \bar{L}_K = \bar{l}_K]}{E[Y\mid \bar{A}_K =  \bm 0, \bar{L}_K = (l_0,\bm 0)]},  \left(\bar{l}_K,  \bar{a}_K \right) \in \{l_0\} \times \left(\{0,1\}^{2K+1} \setminus \{\vec{0}_{2K+1} \}\right); GOP, \bm \eta \right)
		\end{equation*}
		is also an automorphism on 
		$(\mathbb{R}^+)^{d_1} \times (0,1)^{d_2}$.
		
	Now fix a value $l_0$.	What is left to show is that given $\bm \eta$,
		\begin{flalign*}
		(E[Y\mid  \bar{A}_K=\bar{a}_K, &\bar{L}_K=\bar{l}_K], \left(\bar{l}_K, \bar{a}_K \right) \in \{l_0\} \times  \{0,1\}^{2K+1})	\rightarrow  \\
		&  \left(\dfrac{E[Y\mid \bar{A}_K = \bar{a}_{K}, \bar{L}_K = \bar{l}_K]}{E[Y\mid \bar{A}_K =  \bm 0, \bar{L}_K = (l_0,\bm 0)]},  \left(\bar{l}_K,  \bar{a}_K \right) \in \{l_0\} \times \left(\{0,1\}^{2K+1} \setminus \{\vec{0}_{2K+1} \}\right); GOP \right)  \numberthis \label{eqn:mapping_general}
		\end{flalign*}
		is a bijection from $(0,1)^{d_1}$ to $(\mathbb{R}^+)^{d_1}$.

		Arguing as in the proof of Theorem \ref{thm:main}, to show that \eqref{eqn:mapping_general} is bijective, 
		let $\bm c$ be a  vector in $(\mathbb{R}^{+})^{d_1}$. We need to show that there is one and only one $\bm p = \left(p_{\bar{a}_K, \bar{l}_K}, \bar{a}_K \in \{0,1\}^{K+1},  \bar{l}_K \in \{l_0\} \times  \{0,1\}^K\right) \in   (0,1)^{d_1}$ such that 
		\begin{equation}
		\label{eqn:mapping2}
		\left(\dfrac{E[Y\mid \bar{A}_K = \bar{a}_{K}, \bar{L}_K = \bar{l}_K]}{E[Y\mid \bar{A}_K =  \bm 0, \bar{L}_K = (l_0,\bm 0)]},  \bar{a}_K \in \{0,1\}^{K+1},  \bar{l}_K \in \{l_0\} \times \{0,1\}^K; GOP \right)   = \bm c.
		\end{equation}	
		Define $\bm k$ and $p_{max}$ in a manner similar to the proof 
		of Theorem \ref{thm:main}.
		Equation \eqref{eqn:mapping2} may equivalently be expressed as:
		\begin{equation*}
		\log(GOP) = \sum\limits_{\bar{a}_K, {l}_1,\ldots,l_K} \log\left(k_{\bar{a}_K, \bar{l}_K} \right) + d_1  \log\left( p_{max} \right) - \sum\limits_{\bar{a}_K, {l}_1,\ldots,l_K} \log(1-k_{\bar{a}_K, \bar{l}_K} p_{max}) = \log(c_{d_1}).
		\end{equation*}
		Let $g(p_{max}) = \sum\limits_{\bar{a}_K, {l}_1,\ldots,l_K} \log\left(k_{\bar{a}_K, \bar{l}_K} \right) + d_1  \log\left( p_{max} \right) - \sum\limits_{\bar{a}_K, {l}_1,\ldots,l_K} \log(1-k_{\bar{a}_K, \bar{l}_K} p_{max}) - \log(c_{d_1})$. Using the same argument as in the proof of Theorem \ref{thm:main}, one may show that $g(p_{max})$ has one and only one solution in the interval (0,1) so that there is one and only one $\bm p = p_{max}\bm k \in (0,1)^{d_1}$  such that \eqref{eqn:mapping2} holds. 
		
		Lastly, since the domain of $\mathcal{M}$, i.e. $(\mathbb{R}^+)^{d_1} \times (0,1)^{d_2}$ is the Cartesian product of the marginal domains of models in $\mathcal{M}$, models in $\mathcal{M}$ are variation independent of each other.
		
		We have hence finished our proof.
	\end{proof}	
	
\section{Computation of  $r_{\text{max}}$ in Step 3 of Algorithm \ref{alg:estimate_prob}  under the Markov assumption}
	\label{sec:r_max}
	
	
	Note that under assumption \eqref{eqn:markov}, for $m>k,$ we have
	\begin{flalign*}
	\eta((\bar{a}_{k-1},1,\vec{0}_{m-k}), (\bar{l}_k, \vec{0}_{m-k})) &= \eta((\bar{a}_{k-1},0,\vec{0}_{m-k}), (\bar{l}_k, \vec{0}_{m-k})), \\
	\phi((\bar{a}_{k-1},1,\vec{0}_{m-k}), (\bar{l}_k, \vec{0}_{m-k})) &= \phi((\bar{a}_{k-1},0,\vec{0}_{m-k}), (\bar{l}_k, \vec{0}_{m-k}))
	\end{flalign*} 
	so that 
	$$\dfrac{	\eta((\bar{a}_{k-1},1,\vec{0}_{m-k}), (\bar{l}_k, \vec{0}_{m-k})) \phi((\bar{a}_{k-1},1,\vec{0}_{m-k}), (\bar{l}_k, \vec{0}_{m-k}))  + 1-\eta((\bar{a}_{k-1},1,\vec{0}_{m-k}), (\bar{l}_k, \vec{0}_{m-k}))  }   
	{\eta((\bar{a}_{k-1}, 0, \vec{0}_{m-k}), (\bar{l}_{k}, \vec{0}_{m-k})) \phi((\bar{a}_{k-1}, 0, \vec{0}_{m-k}), (\bar{l}_{k}, \vec{0}_{m-k}))   + 1-\eta((\bar{a}_{k-1}, 0, \vec{0}_{m-k}), (\bar{l}_{k}, \vec{0}_{m-k})) } = 1.$$
	Consequently, \eqref{eqn:1} implies that for $k < K$,
	\begin{flalign*}
	\MoveEqLeft[0]{\theta_k(\bar{a}_{k-1}, \bar{l}_k)} \\
	&= \footnotesize \prod\limits_{m=k}^{K-1}  \dfrac{	\eta((\bar{a}_{k-1},1,\vec{0}_{m-k}), (\bar{l}_k, \vec{0}_{m-k})) \phi((\bar{a}_{k-1},1,\vec{0}_{m-k}), (\bar{l}_k, \vec{0}_{m-k}))  + 1-\eta((\bar{a}_{k-1},1,\vec{0}_{m-k}), (\bar{l}_k, \vec{0}_{m-k}))  }   
	{\eta((\bar{a}_{k-1}, 0, \vec{0}_{m-k}), (\bar{l}_{k}, \vec{0}_{m-k})) \phi((\bar{a}_{k-1}, 0, \vec{0}_{m-k}), (\bar{l}_{k}, \vec{0}_{m-k}))   + 1-\eta((\bar{a}_{k-1}, 0, \vec{0}_{m-k}), (\bar{l}_{k}, \vec{0}_{m-k})) }    \\
	& \quad\times\dfrac{E[Y\mid \bar{A}_K = (\bar{a}_{k-1},1, \bm 0), \bar{L}_K = (\bar{l}_k,\bm 0)]}{E[Y\mid \bar{A}_K = (\bar{a}_{k-1}, 0, \bm 0), \bar{L}_K = (\bar{l}_k,\bm 0)]} \\
	&=  \dfrac{	\eta((\bar{a}_{k-1},1), \bar{l}_k) \phi((\bar{a}_{k-1},1), \bar{l}_k)  + 1-\eta((\bar{a}_{k-1},1), \bar{l}_k)  }   
	{\eta((\bar{a}_{k-1}, 0), \bar{l}_{k}) \phi((\bar{a}_{k-1}, 0), \bar{l}_{k})   + 1-\eta((\bar{a}_{k-1}, 0), \bar{l}_{k}) }  \times \dfrac{E[Y\mid \bar{A}_K = (\bar{a}_{k-1},1, \bm 0), \bar{L}_K = (\bar{l}_k,\bm 0)]}{E[Y\mid \bar{A}_K = (\bar{a}_{k-1}, 0, \bm 0), \bar{L}_K = (\bar{l}_k,\bm 0)]}
	\end{flalign*}
	so that 
	\begin{flalign*}
	& \dfrac{E[Y\mid \bar{A}_K = (\bar{a}_{k-1},1, \bm 0), \bar{L}_K = (\bar{l}_k,\bm 0)]}{E[Y\mid \bar{A}_K = (\bar{a}_{k-1}, 0, \bm 0), \bar{L}_K = (\bar{l}_k,\bm 0)]} \\ &= \theta_k(\bar{a}_{k-1}, \bar{l}_k) \left/ 
	\dfrac{	\eta((\bar{a}_{k-1},1), \bar{l}_k) \phi((\bar{a}_{k-1},1), \bar{l}_k)  + 1-\eta((\bar{a}_{k-1},1), \bar{l}_k)  }   
	{\eta((\bar{a}_{k-1}, 0), \bar{l}_{k}) \phi((\bar{a}_{k-1}, 0), \bar{l}_{k})   + 1-\eta((\bar{a}_{k-1}, 0), \bar{l}_{k}) }  \right. \\
	&= \theta(l_k) \left/ 
	\dfrac{	\eta(1,l_k) \phi(1)  + 1-\eta(1,l_k)  }   
	{\eta(0,l_k) \phi(0)   + 1-\eta(0,l_k) }  \right. \numberthis\label{eqn:3}
	\end{flalign*}
	which only depends on $l_k$. One can also see that $\dfrac{E[Y\mid \bar{A}_K = (\bar{a}_{K-1},1), \bar{L}_K = \bar{l}_K]}{E[Y\mid \bar{A}_K = (\bar{a}_{K-1}, 0), \bar{L}_K = \bar{l}_K]}$
	only depends on $l_{K}$.
	
	Similarly,  \eqref{eqn:2} implies that for $k < K-1$,
	\begin{flalign*}
\MoveEqLeft[0]{\phi(\bar{a}_k, \bar{l}_k)} \\
	&=  \prod\limits_{m=k}^{K-2}  \dfrac{\eta((\bar{a}_{k},0,\vec{0}_{m-k}), (\bar{l}_{k}, 1, \vec{0}_{m-k})) \phi((\bar{a}_{k},0,\vec{0}_{m-k}), (\bar{l}_{k}, 1, \vec{0}_{m-k}))  + 1-\eta((\bar{a}_{k},0,\vec{0}_{m-k}), (\bar{l}_{k}, 1, \vec{0}_{m-k}))  }   
	{\eta((\bar{a}_{k},0,\vec{0}_{m-k}), (\bar{l}_{k}, 0, \vec{0}_{m-k})) \phi((\bar{a}_{k},0,\vec{0}_{m-k}), (\bar{l}_{k},  0,\vec{0}_{m-k}))   + 1-\eta((\bar{a}_{k},0,\vec{0}_{m-k}), (\bar{l}_{k}, 0, \vec{0}_{m-k}))}   \\
	&\qquad \times \dfrac{E[Y\mid \bar{A}_K = (\bar{a}_k, \bm 0), \bar{L}_K = (\bar{l}_k,1,\bm 0)]}{E[Y\mid \bar{A}_K = (\bar{a}_{k}, \bm 0), \bar{L}_K = (\bar{l}_k, 0, \bm 0)]} \\
	&= \dfrac{\eta((\bar{a}_{k},0), (\bar{l}_{k}, 1)) \phi((\bar{a}_{k},0), (\bar{l}_{k}, 1))  + 1-\eta((\bar{a}_{k},0), (\bar{l}_{k}, 1))  }   
	{\eta((\bar{a}_{k},0), (\bar{l}_{k}, 0)) \phi((\bar{a}_{k},0), (\bar{l}_{k},  0))   + 1-\eta((\bar{a}_{k},0), (\bar{l}_{k}, 0))}  \times  \dfrac{E[Y\mid \bar{A}_K = (\bar{a}_k, \bm 0), \bar{L}_K = (\bar{l}_k,1,\bm 0)]}{E[Y\mid \bar{A}_K = (\bar{a}_{k}, \bm 0), \bar{L}_K = (\bar{l}_k, 0, \bm 0)]}
	\end{flalign*}
	so that
	\begin{flalign*}
	& \ \ \dfrac{E[Y\mid \bar{A}_K = (\bar{a}_k, \bm 0), \bar{L}_K = (\bar{l}_k,1,\bm 0)]}{E[Y\mid \bar{A}_K = (\bar{a}_{k}, \bm 0), \bar{L}_K = (\bar{l}_k, 0, \bm 0)]} \\ &=  \phi(\bar{a}_k, \bar{l}_k)  \left/ \dfrac{\eta((\bar{a}_{k},0), (\bar{l}_{k}, 1)) \phi((\bar{a}_{k},0), (\bar{l}_{k}, 1))  + 1-\eta((\bar{a}_{k},0), (\bar{l}_{k}, 1))  }   
	{\eta((\bar{a}_{k},0), (\bar{l}_{k}, 0)) \phi((\bar{a}_{k},0), (\bar{l}_{k},  0))   + 1-\eta((\bar{a}_{k},0), (\bar{l}_{k}, 0))} \right. \\
	&=  \phi(a_k)  \left/ \dfrac{\eta(0, 1) \phi(0)  + 1-\eta(0, 1)  }   
	{\eta(0, 0) \phi(0)   + 1-\eta(0, 0)} \right. \numberthis\label{eqn:4}
	\end{flalign*}
	which only  depends on $a_k$. One can also see that $\dfrac{E[Y\mid \bar{A}_K = (\bar{a}_{K-1},0), \bar{L}_K = (\bar{l}_{K-1},1)]}{E[Y\mid \bar{A}_K = (\bar{a}_{K-1},0), \bar{L}_K = (\bar{l}_{K-1}, 0)]} $ only depends on $a_{K-1}$.
	
	Given \eqref{eqn:3} and \eqref{eqn:4}, we can now use dynamic programming to find $r_{\text{max}}$. Specifically, define the following functions:
	\begin{flalign*}
	O(a_k) &= \max_{l_{k+1},\ldots, l_K, a_{k+1}, \dots, a_K} \dfrac{E[Y\mid \bar{A}_K = \bar{a}_K, \bar{l}_K = \bar{l}_K]}{E[Y\mid \bar{A}_K = (\bar{a}_{k},   \vec{0}_{K-k}), \bar{l}_K = (\bar{l}_k, \vec{0}_{K-k})]}, \\	
	O(l_k) &= \max_{l_{k+1},\ldots, l_K, a_{k}, \dots, a_K} \dfrac{E[Y\mid \bar{A}_K = \bar{a}_K, \bar{l}_K = \bar{l}_K]}{E[Y\mid \bar{A}_K = (\bar{a}_{k-1},   \vec{0}_{K-k+1}), \bar{l}_K = (\bar{l}_k, \vec{0}_{K-k})]}.
	\end{flalign*}
	These are well-defined as due to \eqref{eqn:3} and \eqref{eqn:4}, $O(a_k)$ and $O(l_k)$ only depend on $a_k$ and $l_k$, respectively. 
	
	The algorithm proceeds by computing $O(a_k)$ and $O(l_k)$ sequentially. First note that 
	$$
	O(l_K) = \max_{a_K} \dfrac{E[Y\mid \bar{A}_K = \bar{a}_K, \bar{l}_K = \bar{l}_K]}{E[Y\mid \bar{A}_K = (\bar{a}_{K-1},  0), \bar{l}_K = l_K]}.
	$$
	Given the value of $O(l_{k+1})$,  we have
	\begin{flalign*}
	O(a_k) &= \max_{l_{k+1},\ldots, l_K, a_{k+1}, \dots, a_K} \dfrac{E[Y\mid \bar{A}_K = \bar{a}_K, \bar{l}_K = \bar{l}_K]}{E[Y\mid \bar{A}_K = (\bar{a}_{k},   \vec{0}_{K-k}), \bar{l}_K = (\bar{l}_k, \vec{0}_{K-k})]} \\	
	&=  \max_{l_{k+1}}  \left\{ \dfrac{E[Y\mid \bar{A}_K = (\bar{a}_{k},   \vec{0}_{K-k}), \bar{l}_K = (\bar{l}_k, l_{k+1}, \vec{0}_{K-k-1})]}{E[Y\mid \bar{A}_K = (\bar{a}_{k},   \vec{0}_{K-k}), \bar{l}_K = (\bar{l}_k, \vec{0}_{K-k})]}     \times    \right. \\
	& \left.         \max_{l_{k+2},\ldots, l_K, a_{k+1}, \dots, a_K} \dfrac{E[Y\mid \bar{A}_K = \bar{a}_K, \bar{l}_K = \bar{l}_K]}{E[Y\mid \bar{A}_K = (\bar{a}_{k},   \vec{0}_{K-k}), \bar{l}_K = (\bar{l}_k, l_{k+1}, \vec{0}_{K-k-1})]}             \right\}    \\	
	&= \max_{l_{k+1}}  \left\{ \dfrac{E[Y\mid \bar{A}_K = (\bar{a}_{k},   \vec{0}_{K-k}), \bar{l}_K = (\bar{l}_k, l_{k+1}, \vec{0}_{K-k-1})]}{E[Y\mid \bar{A}_K = (\bar{a}_{k},   \vec{0}_{K-k}), \bar{l}_K = (\bar{l}_k, 0, \vec{0}_{K-k-1})]}     \times  O(l_{k+1}) \right\}
	\end{flalign*}
	and similarly
	$$
	O(l_k) =  \max_{a_{k}}  \left\{ \dfrac{E[Y\mid \bar{A}_K = (\bar{a}_{k-1}, a_k,  \vec{0}_{K-k}), \bar{l}_K = (\bar{l}_k, \vec{0}_{K-k})]}{E[Y\mid \bar{A}_K = (\bar{a}_{k-1}, 0,  \vec{0}_{K-k}), \bar{l}_K = (\bar{l}_k, \vec{0}_{K-k})]}     \times  O(a_{k}) \right\}.
	$$
	Finally we get that 
	$$
	O(l_0) = \max_{l_{1},\ldots, l_K, a_{0}, \dots, a_K} \dfrac{E[Y\mid \bar{A}_K = \bar{a}_K, \bar{l}_K = \bar{l}_K]}{E[Y\mid \bar{A}_K =    \vec{0}_{K+1}, \bar{l}_K = (l_0,\vec{0}_{K})]} = r_{\text{max}}.
	$$

	%
	%
	%

\section{Extension of Markov assumption to allow for dependence on the previous two time points}
\label{sec:markov-extension}

\begin{equation}
\label{eqn:markov2}
    \theta_k(\bar{a}_{k-1}, \bar{l}_k) = \theta(a_{k-1}, (l_{k-1},l_k)), \ \phi(\bar{a}_k, \bar{l}_k) = \phi((a_{k-1},a_k), l_k), \ \eta(\bar{a}_k, \bar{l}_k) = \eta((a_{k-1}, a_k), (l_{k-1}, l_k)).  
\end{equation}

	Note that under assumption \eqref{eqn:markov2}, for $m>k+1,$ we have
	\begin{flalign*}
	\eta((\bar{a}_{k-1},1,\vec{0}_{m-k}), (\bar{l}_k, \vec{0}_{m-k})) &= \eta((\bar{a}_{k-1},0,\vec{0}_{m-k}), (\bar{l}_k, \vec{0}_{m-k})), \\
	\phi((\bar{a}_{k-1},1,\vec{0}_{m-k}), (\bar{l}_k, \vec{0}_{m-k})) &= \phi((\bar{a}_{k-1},0,\vec{0}_{m-k}), (\bar{l}_k, \vec{0}_{m-k}))
	\end{flalign*} 
	so that 
	$$\dfrac{	\eta((\bar{a}_{k-1},1,\vec{0}_{m-k}), (\bar{l}_k, \vec{0}_{m-k})) \phi((\bar{a}_{k-1},1,\vec{0}_{m-k}), (\bar{l}_k, \vec{0}_{m-k}))  + 1-\eta((\bar{a}_{k-1},1,\vec{0}_{m-k}), (\bar{l}_k, \vec{0}_{m-k}))  }   
	{\eta((\bar{a}_{k-1}, 0, \vec{0}_{m-k}), (\bar{l}_{k}, \vec{0}_{m-k})) \phi((\bar{a}_{k-1}, 0, \vec{0}_{m-k}), (\bar{l}_{k}, \vec{0}_{m-k}))   + 1-\eta((\bar{a}_{k-1}, 0, \vec{0}_{m-k}), (\bar{l}_{k}, \vec{0}_{m-k})) } = 1.$$
	Consequently, \eqref{eqn:1} implies that for $k < K$,
	\begin{flalign*}
\MoveEqLeft[0]{\theta_k(\bar{a}_{k-1}, \bar{l}_k)} \\
	&= \footnotesize \prod\limits_{m=k}^{K-1}  \dfrac{	\eta((\bar{a}_{k-1},1,\vec{0}_{m-k}), (\bar{l}_k, \vec{0}_{m-k})) \phi((\bar{a}_{k-1},1,\vec{0}_{m-k}), (\bar{l}_k, \vec{0}_{m-k}))  + 1-\eta((\bar{a}_{k-1},1,\vec{0}_{m-k}), (\bar{l}_k, \vec{0}_{m-k}))  }   
	{\eta((\bar{a}_{k-1}, 0, \vec{0}_{m-k}), (\bar{l}_{k}, \vec{0}_{m-k})) \phi((\bar{a}_{k-1}, 0, \vec{0}_{m-k}), (\bar{l}_{k}, \vec{0}_{m-k}))   + 1-\eta((\bar{a}_{k-1}, 0, \vec{0}_{m-k}), (\bar{l}_{k}, \vec{0}_{m-k})) }    \\
	&\quad\times \dfrac{E[Y\mid \bar{A}_K = (\bar{a}_{k-1},1, \bm 0), \bar{L}_K = (\bar{l}_k,\bm 0)]}{E[Y\mid \bar{A}_K = (\bar{a}_{k-1}, 0, \bm 0), \bar{L}_K = (\bar{l}_k,\bm 0)]} \\
	&= \dfrac{	\eta((\bar{a}_{k-1},1), \bar{l}_k) \phi((\bar{a}_{k-1},1), \bar{l}_k)  + 1-\eta((\bar{a}_{k-1},1), \bar{l}_k)  }   
	{\eta((\bar{a}_{k-1}, 0), \bar{l}_{k}) \phi((\bar{a}_{k-1}, 0), \bar{l}_{k})   + 1-\eta((\bar{a}_{k-1}, 0), \bar{l}_{k}) }   \times \dfrac{E[Y\mid \bar{A}_K = (\bar{a}_{k-1},1, \bm 0), \bar{L}_K = (\bar{l}_k,\bm 0)]}{E[Y\mid \bar{A}_K = (\bar{a}_{k-1}, 0, \bm 0), \bar{L}_K = (\bar{l}_k,\bm 0)]}  \\
&\quad \times 	 \dfrac{	\eta((\bar{a}_{k-1},1,0), (\bar{l}_k,0)) \phi((\bar{a}_{k-1},1,0), (\bar{l}_k,0))  + 1-\eta((\bar{a}_{k-1},1,0), (\bar{l}_k,0))  }   
	{\eta((\bar{a}_{k-1}, 0,0), (\bar{l}_{k},0)) \phi((\bar{a}_{k-1}, 0,0), (\bar{l}_{k},0))   + 1-\eta((\bar{a}_{k-1}, 0,0), (\bar{l}_{k},0)) }   \\
&= \dfrac{	\eta(({a}_{k-1},1), (l_{k-1},l_k)) \phi(({a}_{k-1},1), {l}_k)  + 1-\eta(({a}_{k-1},1), (l_{k-1},l_k))  }   
	{\eta(({a}_{k-1}, 0), (l_{k-1},l_k)) \phi(({a}_{k-1}, 0), (l_{k-1},l_k))   + 1-\eta(({a}_{k-1}, 0), (l_{k-1},l_k)) }   \\ 
	& \quad\times \dfrac{E[Y\mid \bar{A}_K = (\bar{a}_{k-1},1, \bm 0), \bar{L}_K = (\bar{l}_k,\bm 0)]}{E[Y\mid \bar{A}_K = (\bar{a}_{k-1}, 0, \bm 0), \bar{L}_K = (\bar{l}_k,\bm 0)]} \times	 \dfrac{	\eta((1,0), ({l}_k,0)) \phi((1,0), 0)  + 1-\eta((1,0), ({l}_k,0))  }   
	{\eta((0,0), ({l}_{k},0)) \phi((0,0), 0)   + 1-\eta((0,0), (l_k,0)) } 
	\end{flalign*}
	so that 
	\[
	\dfrac{E[Y\mid \bar{A}_K = (\bar{a}_{k-1},1, \bm 0), \bar{L}_K = (\bar{l}_k,\bm 0)]}{E[Y\mid \bar{A}_K = (\bar{a}_{k-1}, 0, \bm 0), \bar{L}_K = (\bar{l}_k,\bm 0)]} \]
	only depends on $l_{k-1},a_{k-1}, l_k$.
	
One can also see that 
\[
\dfrac{E[Y\mid \bar{A}_K = (\bar{a}_{K-1},1), \bar{L}_K = \bar{l}_K]}{E[Y\mid \bar{A}_K = (\bar{a}_{K-1}, 0), \bar{L}_K = \bar{l}_K]}
\]
	only depends on $l_{K-1}, a_{K-1}, l_{K}$.
	
	Similarly, for $k < K,$ 
	\[
	\dfrac{E[Y\mid \bar{A}_K = (\bar{a}_k, \bm 0), \bar{L}_K = (\bar{l}_k,1,\bm 0)]}{E[Y\mid \bar{A}_K = (\bar{a}_{k}, \bm 0), \bar{L}_K = (\bar{l}_k, 0, \bm 0)]}
	\]
	only depends on $a_{k-1}, l_k, a_k$.

		We can now use dynamic programming to find $r_{\text{max}}$. Specifically, define the following functions:
	\begin{flalign*}
	O(a_{k-1}, l_k, a_k) &= \max_{l_{k+1},\ldots, l_K, a_{k+1}, \dots, a_K} \dfrac{E[Y\mid \bar{A}_K = \bar{a}_K, \bar{l}_K = \bar{l}_K]}{E[Y\mid \bar{A}_K = (\bar{a}_{k},   \vec{0}_{K-k}), \bar{l}_K = (\bar{l}_k, \vec{0}_{K-k})]}, \\	
	O(l_{k-1}, a_{k-1}, l_k) &= \max_{l_{k+1},\ldots, l_K, a_{k}, \dots, a_K} \dfrac{E[Y\mid \bar{A}_K = \bar{a}_K, \bar{l}_K = \bar{l}_K]}{E[Y\mid \bar{A}_K = (\bar{a}_{k-1},   \vec{0}_{K-k+1}), \bar{l}_K = (\bar{l}_k, \vec{0}_{K-k})]}.
	\end{flalign*}

	The algorithm proceeds by computing $O(a_{k-1}, l_k, a_k)$ and $O(l_{k-1}, a_{k-1}, l_k)$ sequentially. First note that 
	$$
	O(l_{K-1}, a_{K-1}, l_K) = \max_{a_K} \dfrac{E[Y\mid \bar{A}_K = \bar{a}_K, \bar{l}_K = \bar{l}_K]}{E[Y\mid \bar{A}_K = (\bar{a}_{K-1},  0), \bar{l}_K = l_K]}.
	$$
	Given the value of $O(l_k, a_{k+1}, l_{k+1})$,  we have
	\begin{flalign*}
	O(a_{k-1}, l_k, a_k) &= \max_{l_{k+1},\ldots, l_K, a_{k+1}, \dots, a_K} \dfrac{E[Y\mid \bar{A}_K = \bar{a}_K, \bar{l}_K = \bar{l}_K]}{E[Y\mid \bar{A}_K = (\bar{a}_{k},   \vec{0}_{K-k}), \bar{l}_K = (\bar{l}_k, \vec{0}_{K-k})]} \\	
	&=  \max_{l_{k+1}}  \left\{ \dfrac{E[Y\mid \bar{A}_K = (\bar{a}_{k},   \vec{0}_{K-k}), \bar{l}_K = (\bar{l}_k, l_{k+1}, \vec{0}_{K-k-1})]}{E[Y\mid \bar{A}_K = (\bar{a}_{k},   \vec{0}_{K-k}), \bar{l}_K = (\bar{l}_k, \vec{0}_{K-k})]}     \times    \right. \\
	& \left.         \max_{l_{k+2},\ldots, l_K, a_{k+1}, \dots, a_K} \dfrac{E[Y\mid \bar{A}_K = \bar{a}_K, \bar{l}_K = \bar{l}_K]}{E[Y\mid \bar{A}_K = (\bar{a}_{k},   \vec{0}_{K-k}), \bar{l}_K = (\bar{l}_k, l_{k+1}, \vec{0}_{K-k-1})]}             \right\}    \\	
	&= \max_{l_{k+1}}  \left\{ \dfrac{E[Y\mid \bar{A}_K = (\bar{a}_{k},   \vec{0}_{K-k}), \bar{l}_K = (\bar{l}_k, l_{k+1}, \vec{0}_{K-k-1})]}{E[Y\mid \bar{A}_K = (\bar{a}_{k},   \vec{0}_{K-k}), \bar{l}_K = (\bar{l}_k, 0, \vec{0}_{K-k-1})]}     \times  O(l_k, a_{k+1}, l_{k+1}) \right\}
	\end{flalign*}
	and similarly
	$$
	O(l_{k-1}, a_{k-1}, l_k) =  \max_{a_{k}}  \left\{ \dfrac{E[Y\mid \bar{A}_K = (\bar{a}_{k-1}, a_k,  \vec{0}_{K-k}), \bar{l}_K = (\bar{l}_k, \vec{0}_{K-k})]}{E[Y\mid \bar{A}_K = (\bar{a}_{k-1}, 0,  \vec{0}_{K-k}), \bar{l}_K = (\bar{l}_k, \vec{0}_{K-k})]}     \times O(a_{k-1}, l_k, a_k) \right\}.
	$$
	Finally we get that 
	$$
	O(l_{-1}, a_{-1}, l_0) = \max_{l_{1},\ldots, l_K, a_{0}, \dots, a_K} \dfrac{E[Y\mid \bar{A}_K = \bar{a}_K, \bar{l}_K = \bar{l}_K]}{E[Y\mid \bar{A}_K =    \vec{0}_{K+1}, \bar{l}_K = (l_0,\vec{0}_{K})]} = r_{\text{max}},
	$$
	where $l_{-1} = a_{-1} = \emptyset.$

\section{Details on doubly robust estimation}
\label{sec:appendix-dr}

\subsection{Optimal weighting function for the case with two time points}
\label{sec:dr-simu}

According to  \citet[][p.723]{vansteelandt2014structural}, when the variance of $U_k(\alpha)$ given $\bar{L}_k, \bar{A}_k$ is constant, the optimal choice of $d_k (\bar{L}_k, \bar{A}_k)$ is given by 
\begin{align*}
    d_{k, opt}(\Bar{L}_k, \Bar{A}_k) = E[ \dfrac{\partial U_k(\alpha)}{\partial \alpha} | \Bar{L}_k, \Bar{A}_k], k=0,1;
\end{align*}
see also \citet[][p. 212, eqn. (3.11) and (3.12)]{robins2004optimal}.

\subsubsection{Optimal weighting function for $k=1$}

When $k=1,$ following equations \eqref{eqn:simu3} and \eqref{eqn:simu5}, we have 
$$
    U_1(\alpha) = Y \exp(-\alpha_{A_0, L_1}^T L_0 A_1)
$$
so that 
$\hat{d}_{1,opt}^{(0)}(\bar{L}_1,\bar{A}_1) \equiv \hat{E}[ \dfrac{\partial U_1(\alpha)}{\partial \alpha_0} | L_0, A_0, L_1, A_1] = 0.$ and 
\begin{equation}
\label{eqn:partial}
   \hat{d}_{1,opt}^{(a_0,l_1)}(\bar{L}_1,\bar{A}_1) \equiv  \hat{E}[ \dfrac{\partial U_1(\alpha)}{\partial \alpha_{(a_0,l_1)}} | L_0, A_0, L_1, A_1] = - I(A_0=a_0, L_1=l_1)
         L_0 A_1  \hat{E}[Y\mid L_0, A_0, L_1, A_1=0].
\end{equation}
It follows that $\hat{d}_{1,opt} - \hat{E}({d}_{1,opt}\mid L_0, A_0, L_1)$ is a vector with the $(A_0, L_1)$-th element being 
        $- L_0 (A_1-P(A_1=1\mid L_0, A_0,L_1; \hat{\epsilon}))  \hat{E}[Y\mid L_0, A_0, L_1, A_1=0]$ and the rest being zero.

\subsubsection{Optimal weighting function for $k=0$}

When $k=0,$ following equations \eqref{eqn:simu3} and \eqref{eqn:simu5}, we have 
$$
    U_0(\alpha) = Y \exp(-\alpha_{A_0, L_1}^T L_0 A_1 - \alpha_0^T L_0 A_0)
$$
so that 
    \begin{flalign*}
       d_{0,opt}^{(0)}(L_0,A_0) \equiv & E[\dfrac{\partial U_0(\alpha)}{\partial \alpha_0}  | L_0, A_0] = E[-L_0 A_0 U_0(\alpha) \mid L_0, A_0]
 &= -L_0 A_0E_{L_1\mid L_0, A_0}\left[ E\{U_0(\alpha)|L_0, A_0, L_1\}\right] \\
&= -L_0 A_0\exp(-\alpha_0^T L_0 A_0) E_{L_1\mid L_0, A_0}\left[ E\{Y \left(\dfrac{E[Y(A_0,1)\mid L_0, A_0, L_1]}{E[Y(A_0,0)\mid L_0, A_0, L_1]} \right)^{-A_1}  |L_0, A_0, L_1\}\right] \\
&=  -L_0 A_0\exp(-\alpha_0^T L_0 A_0) E_{L_1\mid L_0, A_0} E[Y(A_0,0)\mid L_0, A_0, L_1] \\
&= -L_0 A_0\left(\dfrac{E[Y(1,0)\mid L_0]}{E[Y(0,0)\mid L_0]} \right)^{-A_0} E[Y(A_0,0)\mid L_0, A_0] \\
&= -L_0 A_0E[Y(0,0)\mid L_0]. 
\end{flalign*}
As a result,
\begin{flalign*}
{d}_{0,opt}^{(0)}(L_0,A_0) - E\{{d}_{0,opt}^{(0)}(L_0,A_0)\mid L_0\}  &= -L_0 E[Y(0,0)\mid L_0] \{ A_0 - P(A_0=1\mid L_0)\}, 
\end{flalign*}
in which $E[Y(0,0)\mid L_0]$ may be estimated using equation \eqref{eqn:22}. 

Finally, 
    \begin{align*}
     \dfrac{\partial U_0(\alpha)}{\partial \alpha_{(a_0,l_1)}} &= -L_0 A_1 U_0(\alpha) I(A_0=a_0) I(L_1=l_1) \\
      d_{0,opt}^{(a_0,l_1)}(L_0,A_0)
      &\equiv E[\dfrac{\partial U_0(\alpha)}{\partial \alpha_{(a_0,l_1)}}| L_0, A_0]= -L_0 I(A_0 = a_0) E[A_1 U_0(\alpha)I(L_1=l_1)|L_0, A_0]\\
       &= -L_0 I(A_0 = a_0) \exp(-\alpha_0^T L_0 A_0) E[E[A_1 U_1(\alpha) |\bar L_1,A_0] I(L_1=l_1) |L_0,A_0]\\
       &= -L_0 I(A_0 = a_0) \exp(-\alpha_0^T L_0 A_0)  P(L_1 = l_1|L_0, A_0)  P(A_1=1\mid L_0, A_0, L_1=l_1) \times \\
       &E[Y\mid L_0, A_0, L_1 = l_1, A_1=0]
         \numberthis \label{d_0_k} 
    \end{align*}
so that 
\begin{flalign*}
\MoveEqLeft{d_{0,opt}^{(a_0,l_1)}(L_0,A_0) -  E[d_{0,opt}^{(a_0,l_1)}(L_0,A_0)\mid L_0]}\\
    &= -L_0 \left\{ I(A_0=a_0) - P(A_0=a_0\mid L_0) \right\}  \times \\
    &\exp(-\alpha_0^T L_0 a_0)  P(L_1 = l_1|L_0, A_0=a_0)  P(A_1=1\mid L_0, A_0=a_0, L_1=l_1) \times \\
    &E[Y\mid L_0, A_0=a_0, L_1 = l_1, A_1=0].
      \numberthis \label{E[d_0_k|L_k]}\\ 
\end{flalign*}

\subsection{Doubly robust estimation in the general case }
\label{sec:dr-data}

In the general case, we let 
$$
    U_k(\alpha) = Y \prod\limits_{m=k}^K \theta_m (\bar{A}_{m-1}, \bar{L}_m; \alpha)^{-A_m}, k=0,\ldots, K,
$$
and $\hat{\alpha}_{dr}$ solve the following estimating equation 
\citep{robins1994correcting,vansteelandt2014structural}:
\begin{flalign*}
	\mathbb{P}_n \sum_{k=0}^K \left[ d_k(\bar L_k, \bar A_k) - \hat{E}\{ d_k(\bar L_k, \bar A_k) |\bar L_k,\bar A_{k-1} \} \right]
	 \times \{U_k(\alpha) -\hat{E}(U_k(\alpha) |\bar L_k,\bar A_{k-1} ) \}\Big]=0,
\end{flalign*}
where $d_k(\bar L_k, \bar A_k) $ is a measurable function of the same dimension as $\alpha$. 

Furthermore, we have 
\begin{flalign*}
	E[U_K |\bar L_K,\bar A_{K-1} ] 
	&= E[Y(\bar A_{K-1}, 0)|\bar L_K, \bar A_{K-1}] 
\end{flalign*}
and for $k < K,$ if $E[U_{k+1}\mid \bar{L}_{k+1}, \bar{A}_{k}] = E[Y(\bar A_{k}, \bm 0)|\bar L_{k+1}, \bar A_{k}],$ then
\begin{flalign*}
\MoveEqLeft{E[U_k |\bar L_k,\bar A_{k-1} ]}\\
	&=  	E\left[\left. U_{k+1} \times  \theta_k(\bar{A}_{k-1}, \bar{L}_k)^{-A_k} \right|\bar L_k,\bar A_{k-1} \right] \\
	&=  E_{A_k\mid \bar{L}_k, \bar{A}_{k-1}}E_{L_{k+1}\mid \bar{L}_k, \bar{A_k}}  E\left[ \left. U_{k+1} \times  \left(  \dfrac{E[Y(\bar{A}_{k-1},1, \bm 0) \mid \bar{A}_{k-1}, \bar{L}_k=\bar{l}_k]}{E[Y(\bar{A}_{k-1}, 0, \bm 0) \mid \bar{A}_{k-1}, \bar{L}_k=\bar{l}_k]}  \right)^{-A_k} \right|\bar L_{k+1},\bar A_{k} \right] \\
	&=  E_{A_k\mid \bar{L}_k, \bar{A}_{k-1}} \left(  \dfrac{E[Y(\bar{A}_{k-1},1, \bm 0) \mid \bar{A}_{k-1}, \bar{L}_k=\bar{l}_k]}{E[Y(\bar{A}_{k-1}, 0, \bm 0) \mid \bar{A}_{k-1}, \bar{L}_k=\bar{l}_k]}  \right)^{-A_k}E_{L_{k+1}\mid \bar{L}_k, \bar{A_k}} E\left[ \left. Y(\bar{A}_k,\bm 0)   \right|\bar L_{k+1},\bar A_{k} \right] \\
		&=  E_{A_k\mid \bar{L}_k, \bar{A}_{k-1}} \left(  \dfrac{E[Y(\bar{A}_{k-1},1, \bm 0) \mid \bar{A}_{k-1}, \bar{L}_k=\bar{l}_k]}{E[Y(\bar{A}_{k-1}, 0, \bm 0) \mid \bar{A}_{k-1}, \bar{L}_k=\bar{l}_k]}  \right)^{-A_k} E\left[ \left. Y(\bar{A}_k,\bm 0)   \right|\bar L_{k},\bar A_{k} \right] \\
		&= E[Y(\bar A_{k-1}, \bm 0)|\bar L_{k}, \bar A_{k-1}].
\end{flalign*}



By induction, we have that for $k=0,\ldots, K$
\begin{flalign*}
\MoveEqLeft{E[U_k |\bar L_k,\bar A_{k-1} ]} \\
    	&= E[Y(\bar A_{k-1}, \bm 0)|\bar L_{k}, \bar A_{k-1}] \\
    	&= E[Y(\bar A_{k-1}, \bm 0)|\bar L_{k}, \bar A_{k-1}, A_k = 0] \\
    	&= \eta((\bar{A}_{k-1},0), \bar{L}_k)E[Y(\bar A_{k-1}, \bm 0)|\bar L_{k}, \bar{L}_{k+1} = 1, \bar A_{k-1}, A_k = 0]    
    	\\
    	& \quad +\left\{1-\eta((\bar{A}_{k-1},0), \bar{L}_k)\right\}E[Y(\bar A_{k-1}, \bm 0)|\bar L_{k}, \bar{L}_{k+1} = 0, \bar A_{k-1}, A_k = 0] \\
    	&= \eta((\bar{A}_{k-1},0), \bar{L}_k)E[U_{k+1}\mid (\bar L_{k},1), (\bar A_{k-1}, 0)]  \\
&\quad    	 + \left\{1-\eta((\bar{A}_{k-1},0), \bar{L}_k)\right\}E[U_{k+1}\mid (\bar L_{k},0), (\bar A_{k-1}, 0)] 
\end{flalign*}
may be evaluated recursively.



\section{Additional  Simulations}
\label{sec:addition-simu}

\subsection{Results on computation time}

Table \ref{tab:time} summarizes the computation time for the three proposed methods.

\begin{table}
\centering
\caption{{Mean computation time in seconds for a Monte Carlo sample with size 1000. The computation time for the DR method does not include the time to get the preliminary estimates and warm starting value using 2-step MLE}}
\label{tab:time}
\begin{tabular}{rrrr}
  \toprule
 & MLE & 2-Step MLE & DR \\ 
  \midrule
binary $L_0$ & 29.31 & 26.65 & 4.44 \\ 
  continuous $L_0$ & 566.33 & 514.17 & 130.99 \\
   \bottomrule
\end{tabular}
\end{table}

\subsection{Additional simulation results with continuous $L_0$}

Table \ref{tab:est-L0cont} summarizes the simulation results with continuous $L_0.$

	\begin{table}
		\begin{center}
			\caption{{Bias $\times$ 100 (Monte Carlo standard error $\times$ 100) of the  proposed methods with a continuous baseline covariate $L_0$. 
				The sample size is  1000}}
			\bigskip
			\label{tab:est-L0cont}
			\begin{tabular}{rccccccccccc}
				\toprule
				&       \multicolumn{2}{c}	{MLE} &  \multicolumn{2}{c}	{2-step MLE} &  \multicolumn{2}{c}	{DR}   \\
				\cmidrule(r){2-3} \cmidrule(l){4-5} \cmidrule(l){6-7} 
				& \multicolumn{1}{c}{baseline} & \multicolumn{1}{c}{slope} &   \multicolumn{1}{c}{baseline} & \multicolumn{1}{c}{slope} &   \multicolumn{1}{c}{baseline} & \multicolumn{1}{c}{slope}\\
				\midrule
				\multicolumn{2}{l}{SNMM parameters}	 & & & \\
				$\theta_0(l_0)$ & -0.84(0.45) & -0.14(0.47) & -0.84(0.45) & -0.18(0.47) & 0.61(0.58) & 1.1(0.86) \\ 
  $\theta_1(l_0,1,1)$ & -1.5(1.2) & 1.5(1.1) & -1.5(1.2) & 1.5(1.1) & 1.3(1.5) & 7.7(2.7) \\ 
  $\theta_1(l_0,1,0)$ & -0.21(0.48) & 1.9(0.53) & -0.18(0.48) & 1.9(0.53) & -0.50(0.50) & 2.1(0.73) \\ 
  $\theta_1(l_0,0,1)$ & -1.7(1.1) & 1.1(1.2) & -1.8(1.1) & 1.1(1.2) & 8.0(3.0) & 16(4.2) \\ 
  $\theta_1(l_0,0,0)$ & -0.90(0.54) & 0.35(0.56) & -0.89(0.54) & 0.29(0.56) & 0.023(0.64) & 1.8(0.81) \\   [10pt]
				\multicolumn{2}{l}{Nuisance parameters}	 & & & \\
				  $\phi_0(l_0)$ & -0.29(0.86) & 1.2(0.95) & -0.27(0.86) & 1.1(0.94) & $-$ & $-$ \\ 
  $\phi_1(l_0)$ & -0.093(0.77) & 1.7(0.87) & -0.072(0.77) & 1.8(0.87) & $-$ & $-$ \\ 
  $\text{gop}(l_0)$ & 11(4.0) & 9.7(6.5) & 11(4.0) & 12(6.5) & $-$ & $-$ \\ 
  $\eta_0(l_0)$ & -0.20(0.44) & 0.18(0.39) & -0.18(0.44) & 0.18(0.39) & $-$ & $-$ \\ 
  $\eta_1(l_0)$ & 0.11(0.42) & 0.26(0.37) & 0.061(0.42) & 0.22(0.37) & $-$ & $-$ \\     [10pt]
				\multicolumn{2}{l}{Marginal causal parameters}	 & & & \\[2pt]
				    $E[Y(0,0)]$ & 0.19(0.13) & $-$ & 0.18(0.13) & $-$ & -0.71(0.18) &  $-$ \\[3pt] 
  $E[Y(1,1)]$ & -0.099(0.13) & $-$ & -0.090(0.13) & $-$ & -2.2(0.29) &  $-$ \\[5pt] 
  {\small $\dfrac{E[Y(1,1)]}{E[Y(0,0)]}$} & -0.21(0.39) & $-$ & -0.17(0.39) & $-$ & -2.8(0.66) &  $-$ \\ 
				\bottomrule 
			\end{tabular}
		\end{center}
	\end{table}

\subsection{Illustration of the g-null paradox }
	\label{sec:g-null}


In this part, we illustrate the g-null paradox. Our data are generated following models \eqref{eqn:simu5} -- \eqref{eqn:simu2} and \eqref{eqn:simu1} -- \eqref{eqn:simu4}, 
	where $\epsilon_1 = \epsilon_2 =  (0.1, -0.5)^T, \epsilon_3 = 0.1, \epsilon_4 = -0.5,
	\gamma_0 = (-0.5,-2),  \gamma_1 = (-0.5, 1.5)^T, \alpha_j = (0,0)^T \text{for } j = 0, (0,0), (0,1), (1,0), (1,1), \beta_0 = (-0.5, 1.5)^T, \beta_1 = (-0.5, -1)^T$ and $\delta = (-0.5,0.1)^T.$ Note that under this setting,   the mean potential outcome $E[Y(a_0, a_1)]$ is independent of $(a_0, a_1)$, while  both $E[Y|L_0, A_0, L_1,A_1]$ and $E[L_1|L_0, A_0]$ 	depend on $A_0$.
	We run 500 Monte-Carlo replications of $n=5000$ units. 

	We first estimate $E[Y(a_0, a_1)]$ using the g-computation method, where we assume logistic regression models with main effect terms for
	$E[Y|L_0, A_0, L_1, A_1]$ and $P(L_1 = l_1|L_0, A_0).$ Table \ref{tab:g-null-true}  shows the estimates for log causal relative risks, whose true values are 0. Consistent with the g-null paradox, the point estimates using the g-computation method are large compared to their standard errors.

\begin{table}
\begin{center}
\caption{Mean$\times 100$ (Monte Carlo standard error $\times$ 100) of log causal relative risks obtained using the g-computation method. The sample size is 5000}
			\bigskip
			\label{tab:g-null-true}
\begin{tabular}{rc}
\toprule
Log causal relative risk & Mean$\times 100$ (SE$\times 100$) \\ [5pt]
\midrule
$\log E[Y(1,0)]/E[Y(0,0)]$ & $-$1.07(0.16) \\ [5pt]
  $\log E[Y(1,1)]/E[Y(1,0)]$ & 0.26(0.14) \\[5pt] 
  $\log E[Y(0,1)]/E[Y(0,0)]$ & 0.28(0.14) \\ [5pt]
  $\log E[Y(1,1)]/E[Y(0,0)]$ & $-$0.81(0.23) \\ 
  \bottomrule
\end{tabular}
\end{center}
\end{table}
	
	

We also apply our proposed SNMMs, where we assume that the gop model is misspecified so that $\text{gop}(L_0) = exp(\delta^T  \widetilde{L}_0)$.
Table \ref{tab:g-null} summarizes the results. As expected, our proposed doubly robust estimator has very small bias compared to the standard errors. Although in theory, our 2-step MLE is biased in this case, we still observe small bias for the 2-step MLE under this simulation setup.

\begin{table}
\begin{center}
\caption{{Bias $\times$ 100 (Monte Carlo standard error $\times$ 100) of the two-step MLE and doubly robust estimation (DR) under the g-null. The sample size is  5000}}
			\bigskip
			\label{tab:g-null}
\begin{tabular}{rccccccccc}
\toprule
				&       \multicolumn{2}{c}	{2-step MLE} &  \multicolumn{2}{c}	{DR}   \\
				\cmidrule(r){2-3} \cmidrule(l){4-5} 
				& \multicolumn{1}{c}{baseline} & \multicolumn{1}{c}{slope} &   \multicolumn{1}{c}{baseline} & \multicolumn{1}{c}{slope} \\
				\midrule
				\multicolumn{1}{l}{SNMM parameters }	 & & & \\
$\theta_0(l_0)$ & -0.16(0.25) & 0.47(0.40) & -0.12(0.25) & 0.54(0.40) \\ 
  $\theta_1(l_0,1,1)$ & -0.087(0.53) & -0.29(0.99) & -0.14(0.54) & 0.099(1.0) \\ 
  $\theta_1(l_0,1,0)$ & 0.40(0.26) & -0.97(0.42) & 0.36(0.26) & -0.58(0.43) \\ 
  $\theta_1(l_0,0,1)$ & -0.52(0.55) & 0.87(0.71) & -0.44(0.56) & 0.83(0.70) \\ 
  $\theta_1(l_0,0,0)$ & 0.24(0.27) & -0.46(0.52) & 0.24(0.27) & -0.29(0.52) \\ 
   \bottomrule
\end{tabular}
\end{center}
\end{table}

\end{document}